\documentclass[journal,10pt]{IEEEtran}
\IEEEoverridecommandlockouts
\usepackage{cite}
\usepackage{comment}
\usepackage{amsmath,amssymb,amsfonts}
\usepackage{blindtext,graphicx}
\usepackage{textcomp}
\usepackage{epsfig}
\usepackage{caption}
\usepackage{subcaption}
\usepackage{booktabs,makecell}
\usepackage{booktabs}
\usepackage{makeidx}
\usepackage{algorithm}
\usepackage[noend]{algpseudocode}
\usepackage{amsthm}
\usepackage{tikz}
\usepackage{adjustbox}
\usepackage{flafter}
\usepackage{color}
\usepackage{algcompatible}
\usepackage{arydshln}
\usepackage{multirow}
\usepackage{dsfont}
\usepackage{enumitem}

\usepackage{algorithm}
\usepackage[noend]{algpseudocode}
\usepackage{amsthm}

\usepackage{color}

\date{}
\def\BibTeX{{\rm B\kern-.05em{\sc i\kern-.025em b}\kern-.08em
		T\kern-.1667em\lower.7ex\hbox{E}\kern-.125emX}}
\begin{document}
	
	\title{\huge{ Deep Learning-Based Auto-Encoder for Time-Offset Faster-than-Nyquist Downlink NOMA with Timing Errors and Imperfect CSI}}	
	\author{{Ahmed Aboutaleb, Mohammad Torabi, Benjamin Belzer, and Krishnamoorthy Sivakumar}
		\thanks{The authors are with the School of EECS, Washington State University, Pullman, WA 99164, USA. Corresponding author: Benjamin Belzer; Email: belzer@wsu.edu. This work was supported in part by the US NSF under Grant CCF-1817083.}}
	\maketitle
	\vspace{-18mm}
	\begin{abstract}
We examine encoding and decoding of transmitted sequences for the downlink time-offset faster than Nyquist signaling non-orthogonal multiple access NOMA (T-NOMA) channel. We employ a previously proposed  singular value decomposition (SVD)-based scheme as a benchmark. While this SVD scheme provides reliable communication, our findings reveal that it is not optimal in terms of bit error rate (BER). Additionally, the SVD is sensitive to timing offset errors, and its time complexity increases quadratically with the sequence length. We propose a convolutional neural network (CNN) auto-encoder (AE) for encoding and decoding with linear time complexity. We explain the design of the encoder and decoder architectures and the training criteria. By examining several variants of the CNN AE, we show that it can achieve an excellent trade-off between performance and complexity. The proposed CNN AE surpasses the SVD method by approximately 2 dB in a T-NOMA system with no timing offset errors or channel state information estimation errors. In the presence of channel state information (CSI) error variance of 1$\%$ and uniform timing error at $\pm$4\% of the symbol interval, the proposed CNN AE provides up to 10 dB SNR gain over the SVD method. We also propose a novel modified training objective function consisting of a linear combination of the traditionally used cross-entropy (CE) loss function and a closed-form expression for the bit error rate (BER) called the Q-loss function. Simulations show that the modified loss function achieves SNR gains of up to 1 dB over the CE loss function alone. Finally, we investigate several novel CNN architectures for both the encoder and decoder components of the AE that employ additional linear feed-forward connections between the CNN stages; experiments show that these architectural innovations achieve additional SNR gains of up to 2.2 dB over the standard serial CNN AE architecture. 
	\end{abstract}
	
	\begin{IEEEkeywords}
		Asynchronous transmission, auto-encoder, deep learning, faster-than Nyquist signaling, neural networks, non-orthogonal multiple access.
	\end{IEEEkeywords}
\section{Introduction}	

Next-generation wireless standards will support more users, reduce latencies, facilitate high data rates, and improve reliability \cite{NOMAforCellularFutureRadioAccess_Saito}. Current wireless standards employ orthogonal multiple access (OMA) techniques, where distinct users can only communicate over orthogonal resources. An example is orthogonal frequency-division multiple access (OFDMA), which assigns orthogonal sub-carriers to different users to limit inter-user interference (IUI) \cite{OFDMA_A_Broadband_wireless_access_techology_Yin}. However, OMA techniques are sub-optimal in terms of achieving the channel capacity \cite{FasterthanNyquistBroadcastinginGaussianChannelsAchievableRateRegionsandCoding_Kim}.
With non-orthogonal multiple access (NOMA), multiple users can communicate over the same resource element (frequency sub-carriers, time slots, or spreading codes) \cite{NOMAforCellularFutureRadioAccess_Saito,ASurveyofNOMAfor5GNetworksResearchChallengesandFutureTrends_Ding}. 

\subsection{Downlink NOMA}
Consider a downlink channel where a base station communicates to two users using the same orthogonal resource. In downlink NOMA, the channel capacity can be achieved by superposition encoding along with successive interference cancellation (SIC). The base station transmits the superposition of two non-orthogonal signal components. Each component is intended for a particular recipient. Assume that User 1 has better channel conditions than User 2. To recover their own message, User 1 decodes the message of User 2 first. Ideally, at User 1, User 2's message can be decoded correctly and removed from the superposition due to User 1's better channel conditions. Then, User 1 can decode their respective message without interference. User 2 decodes their own message with interference from User 1's message, which cannot be canceled due to User 2's worse channel conditions \cite[pp. 509--595]{Info_Theory_Cover}. We refer to the mentioned technique as stronger/weaker user selection. Power-domain NOMA (P-NOMA) relies mainly on allocating different powers to different users \cite{TimeAsynchronousNOMAforDownlinkTransmission_Ganji,NOMAforCellularFutureRadioAccess_Saito, Concept_and_Practical_Considerations_of_NOMA_for_Future_Radio_Access_Benjebbour}.  

\subsection{Faster than Nyquist Signaling}

In existing wireless standards, the transmission pulse satisfies the Nyquist intersymbol interference (ISI) criterion. The time-domain Nyquist criterion for a pulse $h(t)$ is that $h(0) = 1$ and $h(nT) = 0$, where $T$ is the symbol interval and $n$ is an integer\cite{FTN_Anderson}. If the noise-free pulse shape after matched filtering satisfies the Nyquist criterion, then by sampling the matched filter output at the Nyquist data transmission rate, the received signal does not experience inter-symbol interference (ISI). Hence, maximum likelihood (ML) symbol detection can be achieved using symbol-by-symbol processing without considering channel memory, resulting in simpler implementations. If data is transmitted at a rate faster than the Nyquist rate, then the received signal experiences ISI, and symbol-by-symbol processing no longer provides ML detection performance. However, with appropriate data processing to account for this ISI, faster than Nyquist (FTN) signaling enables higher Shannon capacities, improved bandwidth efficiencies, and multiple user broadcasting \cite{FTN_Anderson,FasterthanNyquistbraodcastsignaling_Kim}. 

\subsection{Time-offset with faster than Nyquist Signaling NOMA (T-NOMA)}
Time-offset with faster than Nyquist signaling NOMA (T-NOMA) introduces timing offsets between the continuous-time signals intended for different users and transmits the superposition of such signals \cite{FasterthanNyquistbraodcastsignaling_Kim,TimeAsynchronousNOMAforDownlinkTransmission_Ganji,AsynchronousTrainsmissionforMACsRateRegionAnalysisandSystemDesignforUplinkNOMA_Ganji}. This technique is referred to as sub-FTN-NOMA in \cite{FasterthanNyquistbraodcastsignaling_Kim}. For decoding in T-NOMA, the received time-domain signal is sampled at a faster rate than the Nyquist transmission rate.
T-NOMA leverages FTN signaling for the broadcast channel, as proposed by Kim and Bajcsy in \cite{FasterthanNyquistbraodcastsignaling_Kim}. 
FTN broadcasting achieves the channel capacity of the broadcast channel.
Furthermore, maximizing the sum rate in T-NOMA facilitates fairness among users, contrary to P-NOMA  \cite{TimeAsynchronousNOMAforDownlinkTransmission_Ganji}.


\subsection{Auto-encoder}
The auto-encoder (AE) was first proposed in \cite{NonlinearPCAAnslysisUsingAutoassociativeNeuralNetworks_Kramer} as a non-linear generalization of principal component analysis. The AE consists of a neural network-based encoder and decoder that are trained jointly to minimize an objective function. For example, in image reconstruction, the encoder learns efficient representations of the input images and the decoder learns to recover the images from such representations. The mean squared error (MSE) loss function is commonly used for training the AE. 
The work in \cite{AnIntroductiontoDeepLearningforthePysicalLayer_OShea} by O'Shea and Hoydis is among the first to illustrate the application of AEs to wireless channels, including a two-user interference channel and a single user fading channel. The AE described in \cite{AnIntroductiontoDeepLearningforthePysicalLayer_OShea} accepts the messages from distinct users for transmission. The encoder encodes the messages such that the decoder can reliably decode the transmitted messages.

\subsection{Relevant Studies on Auto-encoders for NOMA}

A major challenge in T-NOMA is the design of efficient encoding and decoding schemes for data recovery. A linear singular value decomposition (SVD) based encoding and decoding scheme is proposed in \cite{TimeAsynchronousNOMAforDownlinkTransmission_Ganji,AsynchronousNOMAforDownlinkTransmissions_Cui,Power_Domain_Multiplexed_Precoded_Faster_Than_Nyquist_Signaling_for_NOMA_Downlink_Chaki} for T-NOMA. In \cite{TimeAsynchronousNOMAforDownlinkTransmission_Ganji,AsynchronousNOMAforDownlinkTransmissions_Cui}, the authors consider achievable rates for P-NOMA and the SVD scheme in T-NOMA. The SVD scheme enables the receiver to decouple the superposition signal sequence into independent channels with minimal interference assuming accurate knowledge of the transmitters' timing offsets. However, the SVD's encoding and decoding complexities are \emph{quadratic} in the length of the transmitted sequence. In \cite{Power_Domain_Multiplexed_Precoded_Faster_Than_Nyquist_Signaling_for_NOMA_Downlink_Chaki}, Chaki and Sugiura propose a linear encoding and decoding scheme for T-NOMA based on the SVD. Their results show that the proposed scheme achieves competitive BERs and spectral efficiencies using the raised cosine pulse (RCP) compared with the rectangular pulse.
In this paper, we propose a convolutional neural network (CNN) auto-encoder (AE) for T-NOMA whose complexity is \emph{linear} in the length of the transmitted sequence and is more resilient to timing offset errors.

When used with channel coding, SIC for T-NOMA is effective for eliminating the weaker user's interference \cite{FasterthanNyquistbraodcastsignaling_Kim}. But SIC suffers from a detection BER floor if channel coding is not used, because in that case, reliable subtraction of the weaker user's interfering signal cannot be performed. In contrast, the SVD and the AE methods do not inherently suffer from detector BER floors in the absence of channel coding, highlighting their more efficient implementations.

In \cite{DeepLearningforanEffectiveNOMAScheme_Gui}, Gui \emph{et al.} propose an AE consisting of fully connected and long short-term memory (LSTM) layers for P-NOMA. To address imperfect interference cancellation, Kang \emph{et al.} in \cite{DeepLearningBasedMIMONOMAwithImperfectSICDecoding_Kang} propose an AE for encoding and decoding in P-NOMA using fully-connected neural networks (FNNs). In \cite{DeepLearnignBasedJointDetectionandDecodingforNOMASystems_Sun}, Sun \emph{et al.} propose using an FNN for detection and message passing with another FNN decoder in P-NOMA. In \cite{DeepNOMAAUnifiedFrameworkforNOMAUsingDeepMultiTaskLearning_Ye}, Ye \emph{et al.} propose using a deep neural network (DNN) AE consisting of FNNs for P-NOMA. The encoder encodes messages over the available orthogonal resource elements, and the decoder recovers the transmitted messages. In \cite{ANovelMultitaskLearningEmpoweredCodebookDesignforDownlinkSCMANetworks_Luo_2022}, Luo \emph{et al.} propose an AE to learn a sparse code multiple access for downlink NOMA. They show that the AE outperforms previously proposed sparse codes in terms of average BER and complexity. In addition, Jiang \emph{et al.} in \cite{Jiang_A_Residual_Learning_aided_Convolutional_AE_for_SCMA} present a CNN AE with residual connections for sparse code multiple access channels. In \cite{A_Wasserstein_GAN_Autoencoder_for_SCMA_Networks_Miuccio}, Miuccio \emph{et al.} propose integrating a generative adversarial network (GAN) with an AE for sparse code multiple access code design for NOMA. Their results show improved BER performance, especially at low SNRs. In \cite{NOMA_Approaching_Single_User_Bit_Error_Rate_Performance_Han}, Han \emph{et al.} propose a deep AE for sparse code multiple access where both the resource allocation and bit-to-symbol mapping are jointly optimized; their approach achieves BER performance improvements that are competitive with a baseline one-user channel. In \cite{Kang_ML_based_NOMA_for_Multiuser_MISO_Broadcast_Channel}, Kang \emph{et al.} propose using a neural network for finding the user-message decoding order in downlink NOMA with multiple antennas at the base station. In \cite{Parmar_Modulation_Classification_for_NOMA_System_Using_Modified_Residual_CNN}, Parmar \emph{et al.} present a residual CNN for modulation classification in NOMA.

\subsection{Main Contributions}
Despite the recent progress in deep learning systems for NOMA, the proposed approaches assume NOMA with Nyquist rate transmission. When considering T-NOMA, the AE must consider the full length of the transmitted sequence for encoding and decoding. Hence, the AE must be designed such that long sequences are processed efficiently.


The contributions of this paper are summarized as follows.
\begin{enumerate}
	\item We propose novel CNN AE architectures for the T-NOMA downlink channel with linear complexity in the length of the transmitted sequence. To the best of the authors' knowledge, no previous studies have examined the design of AEs for T-NOMA. We show that the proposed AE design outperforms the SVD-based scheme in terms of detection accuracy, robustness, and efficiency. 
	\item We propose a novel objective loss function that combines conventional cross-entropy loss with a Q-function-based term that estimates the error probability using the mean and variance of empirical log-likelihood ratios (LLRs). We show that this modified loss function results in better BERs given the same neural network architecture.
	\item We evaluate the impact of timing error and imperfect channel state information (CSI) on the performance of the proposed and baseline systems. We show that such impairments result in data-dependent noise at the receiver. To improve the CNN AE's robustness to this data-dependent noise, we propose a multilayer perceptron (MLP) power allocator (MLP-PA) at the transmitter and an MLP CSI transformer (MLP-T) at the receiver. We show that the MLP components improve the performance of the CNN AE.  When BPSK modulation is used in a two-user T-NOMA system, the CNN AE system achieves twice the throughput of a single user system at the same BER, given the same average power per user.
    \item For comparison, we also consider a two-user T-NOMA scheme with stronger/weaker users selection under perfect CSI and no timing error. We present a BER comparison between T-NOMA with user selection, T-NOMA with SVD, and T-NOMA with CNN AE architectures. In addition, we derive ergodic achievable rates and the BER for the considered T-NOMA with user selection. We compare those achievable rates with the rates and BERs of T-NOMA with SVD.
\end{enumerate}

\section{System Model}
\subsection{Downlink Channel Model for T-NOMA}
\label{Channel_Model_Subsection}
Let $x_k[n]$ denote the modulation symbol intended for user $k$ at time $n$, $T$ be the symbol interval, $p(t)$ be the truncated pulse shape spanning $N_p$ symbols for the duration $T_p = N_pT$, $N$ be the block length, and $P_{k,n}$ be the transmit power allocated for the $n$th symbol intended for the $k$th user. Then, the complex baseband component of the transmitted signal intended for user $k$ is given by \cite{TimeAsynchronousNOMAforDownlinkTransmission_Ganji}:
\begin{align}
	x_k(t)=\sum_{n=1}^{N}\sqrt{P_{k,n}}x_k[n]p(t-(n-1)T).
\end{align}

Let $K$ denote the number of users, and $\tau_k$ denote the time offset introduced in the signal intended for user $k$. The base station transmits the superposition signal
$x_{\text{transmit}} = \sum_{k=1}^{K}x_k(t-\tau_k). \label{ASynchSignal}$
If $h_r$ is the flat fading channel complex coefficient for the $N$ symbols when communicating to user $r$, then user $r$'s received signal is given by:
\begin{align}
	y^r(t) = h_r\sum_{k=1}^{K}x_k(t-\tau_k)+n^r(t), \label{ContinuousTimeReceivedSignal}
\end{align}
where $n^r(t)$ is complex additive white Gaussian noise (CAWGN) with power spectral density $N_0$ representing the electronics noise at the user equipment.

As shown in \cite{FasterthanNyquistbraodcastsignaling_Kim,TimeAsynchronousNOMAforDownlinkTransmission_Ganji}, sufficient statistics at the receiver for user $r$ for estimating $x_k[n]$ are given by:
\begin{align}
	y_{l}^r[m] &= \int_{-\infty}^{\infty}  y^r(u) p(u - (m-1)T - \tau_l)du 
	\nonumber\\&=(y^r(t)*p(-t))\vert_{t= (m-1)T + \tau_l}, \label{SufficientStatisticsEquation} \\
	&\quad r=1,\ldots,K,~l=1,\ldots,K,~m=1,\ldots,N. \nonumber
\end{align}

Defining $g(t) \triangleq p(t)*p(-t)$ and $g_{l,k}[m-n] \triangleq g((m-n)T + (\tau_l - \tau_k)) $, the discrete-time sufficient statistics are
\begin{align}
	y_l^r[m] = h_r\sum_{k=1}^K \sum_{n=1}^N \sqrt{P_{k,n}}g_{l,k}[m-n]x_k[n] + n_{l}^{r}[m]. \label{Discrete_Time_System_Model_Equation}
\end{align}
We derive correlator receivers for obtaining two-user sufficient statistics with an offset of $T/2$ in \cite[Appendix A].

Without loss of generality, assume $\mathbf{E}\{|x_k[m]|^2\} = 1$, $\mathbf{E}\{|h_r|^2\} = 1$, and $\sum_{m} {g}_{l,k}^2[m] = 1$, for each $l, k$. Let $P_k$ denote the average power allocated to user $k$. Then, $P_k= (1/N)\sum_{n} P_{k,n}$, and the average total power is $P = \sum_k P_k$. The correlated noise $ n_{l}^{r}[m] = n(t)*p(t)|_{t=(m-1)T+\tau_l}$ has auto-correlation:
\begin{align}
	R_l^r[m]= \mathbf{E}[n_{l}^{r}[n+m]n_{l}^{r*}[n]] =
	\begin{cases} N_0g_{l,r}[m], &\text{if $ m \leq N_p$}, \\
		0,  &\text{otherwise.}
	\end{cases}
\end{align} 


\subsection{SVD Baseline}
 When computing the convolution between $x_k$ and $g_{l,k}$ in \eqref{Discrete_Time_System_Model_Equation}, only the $N_v = N-N_p+1$ valid outputs that do not require zero padding are considered. The received signal in \eqref{Discrete_Time_System_Model_Equation} can be written in matrix form as follows. Let $\mathbf{y}^r_l= [y_l^r[1],\ldots, y_l^r[N_v]]^T$, $\mathbf{y}^r = [\mathbf{y}^{rT}_1, \ldots, \mathbf{y}^{rT}_K]^T$, and $\mathbf{n}^r$ be similarly defined. The transmitted symbols are interleaved such that $\mathbf{x} = [\mathbf{x}[1], \ldots, \mathbf{x}[N]]^T$, where $\mathbf{x}[n] = [x_1[n],\ldots,x_K[n]]$ \cite{FasterthanNyquistbraodcastsignaling_Kim}. Then, $\mathbf{y}^r$ is
\begin{align}
	\mathbf{y}^r = h_r\mathbf{G}\mathbf{P}\mathbf{x}+\mathbf{n}^r,
    \label{eq:SVD_yr}
\end{align}
where $\mathbf{G}$ is a $KN_v\times KN$ doubly-block Toeplitz matrix representing the superposition and convolutions using $\mathbf{g}_{l,k}$s, and $\mathbf{P} = \text{diag}([\sqrt{P_{k,n}}])$.

The transmitter can encode the transmitted symbols by premultiplying $\mathbf{Px}$ by a unitary matrix $\mathbf{U}_{\text{transmit}}$. Then, the receiver premultiplies the received signal sequence by the unitary matrix $\mathbf{U}_{\text{receive}}$. Using the appropriate choice of $\mathbf{U}_{\text{transmit}}$ and $\mathbf{U}_{\text{receive}}$, the superposition channel can be decoupled into separate channels. If $\mathbf{G}$ has rank $r_{\mathbf{G}}$, an SVD of $\mathbf{G}$ is given by $
\mathbf{G}=\mathbf{U}\mathbf{\Lambda}\mathbf{V}^H$,
where $\mathbf{U}$ and $\mathbf{V}$ are unitary matrices,
and $\mathbf{\Lambda}$ is a rectangular diagonal matrix with $r_{\mathbf{G}}$ non-zero singular values. If $\mathbf{U}_{\text{transmit}} = \mathbf{V}$ and $\mathbf{U}_{\text{receive}} = \mathbf{U}^H$, then, ideally, the received signal can be decoupled into $r_{\mathbf{G}}$ channels without IUI as follows:
\begin{align}
	\mathbf{U}_{\text{receive}}\mathbf{y}^r
	&= h_r	\mathbf{U}_{\text{receive}}\mathbf{G} \mathbf{U}_{\text{transmit}}\mathbf{P}\mathbf{x} + 	\mathbf{U}_{\text{receive}}\mathbf{n}^r \\ \Leftrightarrow \mathbf{U}^H\mathbf{y}^r  &= h_r\mathbf{\Lambda}\mathbf{P}\mathbf{x} + \mathbf{U}^H\mathbf{n}^r.
	 \label{IndependentChannelsEquation}
\end{align} 
From \eqref{IndependentChannelsEquation}, the diagonal matrix $\mathbf{\Lambda}$ premultiplies $\mathbf{P}\mathbf{x}$, resulting in $r_{\mathbf{G}} \leq KN_v$ decoupled sub-channels seen by the receiver. Ideally, the $r_{\mathbf{G}}$ sub-channels do not contain ISI or IUI. Thus, the detection does not use SIC, which can be difficult to realize in practice. The SVD encoding and decoding applies linear operations on the discrete transmitted and received samples, respectively. The encoding and decoding complexity of the SVD is $O(KN_v)$ as discussed in Subsection \ref{Complexity_comparison_section}.
	
	\subsection{Sources of Data-dependent Noise}
	We illustrate that errors in the timing offsets cause data-dependent noise. Consider the two-user downlink channel where the intended difference between the timing offsets is $\tau_{\text{design}} = \tau_{1,\text{design}} - \tau_{2, \text{design}}$. Without loss of generality, we assume $\tau_{2, \text{design}}=0$. Due to propagation delays, users' mobility, and sample timing errors, the actual time offset is $\tau = \tau_{\text{design}} + \epsilon $, where $\epsilon$ is the timing error. We assume $\epsilon$ is zero-mean and uniformly distributed over a width $w>0$, i.e., $\epsilon \sim U(-w/2,w/2)$.
	Because of the timing error, the transmitted symbols undergo the matrix transformation $\mathbf{G}_{\epsilon}$ instead of $\mathbf{G}$. Let $\mathbf{G}_{\epsilon} = \mathbf{U}_{\epsilon}\mathbf{\Lambda}_{\epsilon}\mathbf{V}_{\epsilon}^H$ be an SVD of $\mathbf{G}_{\epsilon}$. Let $\mathbf{U}^H\mathbf{U}_{\epsilon} = \mathbf{I} + \mathbf{E}_u$ and $\mathbf{V}^H_{\epsilon}\mathbf{V} = \mathbf{I} + \mathbf{E}_v$. Then, the SVD encoding and decoding gives:
	\begin{align}
		\mathbf{U}^H\mathbf{y}^r &= h_r\mathbf{U}^H\mathbf{G}_{\epsilon}\mathbf{V}\mathbf{Px}+\mathbf{U}^H\mathbf{n}^r \nonumber\\ &=h_r(\mathbf{I}+\mathbf{E}_u)\mathbf{\Lambda}_{\epsilon}(\mathbf{I}+\mathbf{E}_v)\mathbf{Px}+\mathbf{U}^H\mathbf{n}^r \nonumber\\
		&= h_r\mathbf{\Lambda}_{\epsilon}\mathbf{Px} + h_r\mathbf{n}_{\epsilon}(\mathbf{x}) + \mathbf{U}^H\mathbf{n}^r, \label{MatrixSystemModelwithTimingErrorEquation} 		
	\end{align} 
	where $\mathbf{n}_{\epsilon}(\mathbf{x}) \triangleq (\mathbf{\Lambda_{\epsilon}}\mathbf{E}_v + \mathbf{E}_u\mathbf{\Lambda}_{\epsilon} + \mathbf{E}_u\mathbf{\Lambda}_{\epsilon}\mathbf{E}_v)\mathbf{Px} $ is the data-dependent noise term due to the timing offset error $\epsilon$.
	
	Another source of data-dependent noise is the estimation error of $h_r$. To correctly detect the transmitted symbols, $h_r$ is usually estimated by the receiver. To enable estimating $h_r$ at the receiver, a pilot symbol is transmitted by the base station to the user. This pilot symbol is known at the receiver, and can be appended to the beginning of the transmitted sequence. The estimated $h_r$, denoted $\hat{h}_r$, is expected to be accurate for the duration $TN$. In some communication systems, the CSI estimate at the receiver is fed back to the transmitter for power allocation. In practice, the estimation of $h_r$ introduces an error $\delta$ such that $\hat{h}_r = h_r + \delta$. Hence, for detection, multiplying the received signal by $\hat{h}_r^*$ gives:
	\begin{align}
		\hat{h}_r^* \mathbf{y}^r &=  (h_r^* + \delta^*)h_r\mathbf{G}\mathbf{P}\mathbf{x} + (h_r^* + \delta^*)\mathbf{n}^r \nonumber\\&= |h_r|^2 \mathbf{G}\mathbf{P}\mathbf{x} +\delta^*h_r\mathbf{G}\mathbf{P}\mathbf{x} + (h_r^* + \delta^*)\mathbf{n}^r. \label{Effect_of_Estimation_hr_ErrorEquation}
	\end{align}
	The term $\delta^*h_r\mathbf{G}\mathbf{P}\mathbf{x}$ in \eqref{Effect_of_Estimation_hr_ErrorEquation} is data-dependent and unknown to the receiver. The error $\delta$ can be modeled as a complex Gaussian random variable with zero mean, i.e., $\delta \sim \mathcal{CN}(0, \sigma^2_{\delta})$.
	 From \eqref{MatrixSystemModelwithTimingErrorEquation} and \eqref{Effect_of_Estimation_hr_ErrorEquation}, the effect of both timing and CSI estimation errors on the received signal is:
	\begin{align}
			&\hat{h}_r^*\mathbf{U}^H\mathbf{y}^r = \nonumber \\ &
			|h_r|^2\mathbf{\Lambda}_{\epsilon}\mathbf{Px} +|h_r|^2\mathbf{n}_{\epsilon}(\mathbf{x})+ \mathbf{n}_{\delta}(\mathbf{x}) + (h_r^* + \delta^*)\mathbf{U}^H\mathbf{n}^r, \label{ImpactTEandImperfectCSIEquation}
	\end{align}
	where $\mathbf{n}_{\delta}(\mathbf{x}) \triangleq \delta^*h_r \mathbf{n}_{\epsilon}(\mathbf{x}) + \delta^*h_r\mathbf{\Lambda}_{\epsilon}\mathbf{Px}$ is the data-dependent noise due to the error on the CSI estimate. Thus, the SVD method is sensitive to timing offset and CSI estimation errors.

	\section{Proposed Auto-Encoder for T-NOMA}
	\begin{figure}[t]
		\centering
		\includegraphics[width=0.49\textwidth]{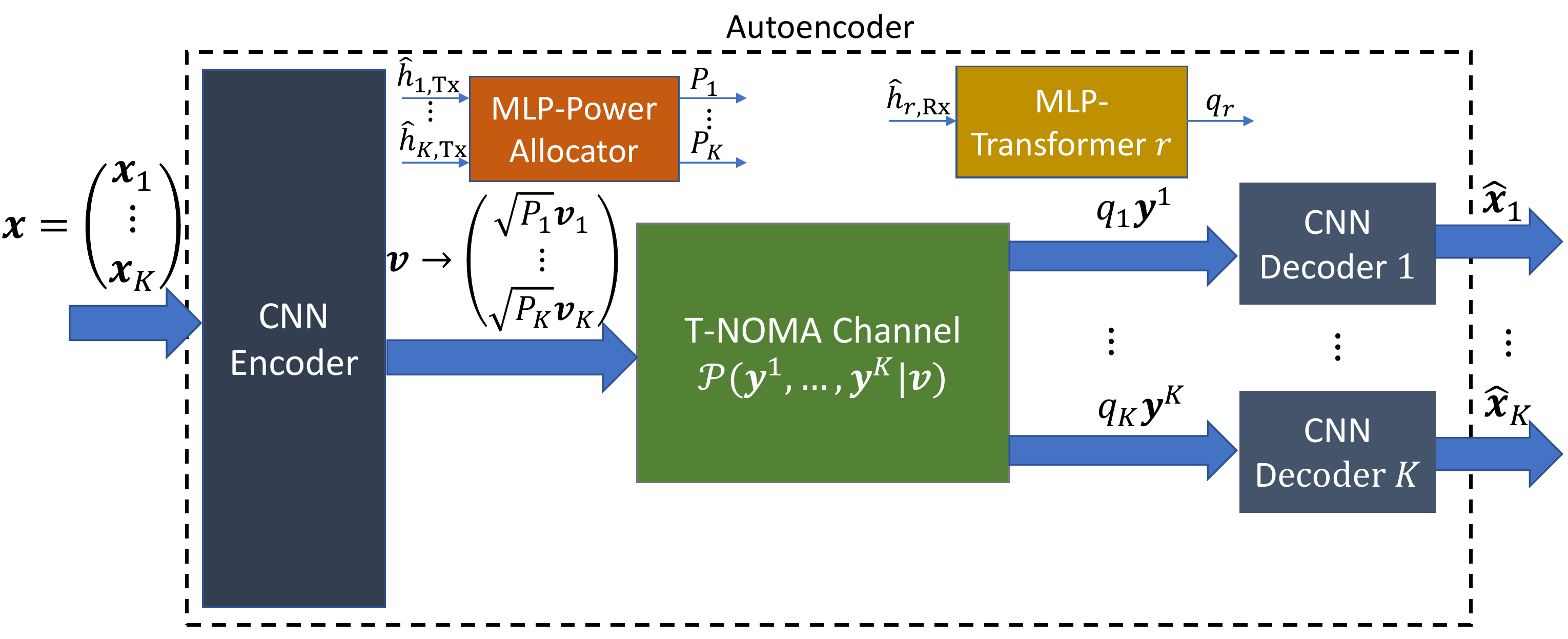}
		\caption{{\small CNN autoencoder  system. The encoder encodes the messages $\mathbf{x}$ for transmission as $\mathbf{v}$. The stochastic T-NOMA channel superimposes the transmitted sequences and uses faster than Nyquist signaling. The decoder at each user recovers the transmitted messages from the received sequence.}}
		\label{CNN_AE_NOMA}
	\end{figure}

	\begin{figure}[t]
		\centering
		\includegraphics[width=0.49\textwidth]{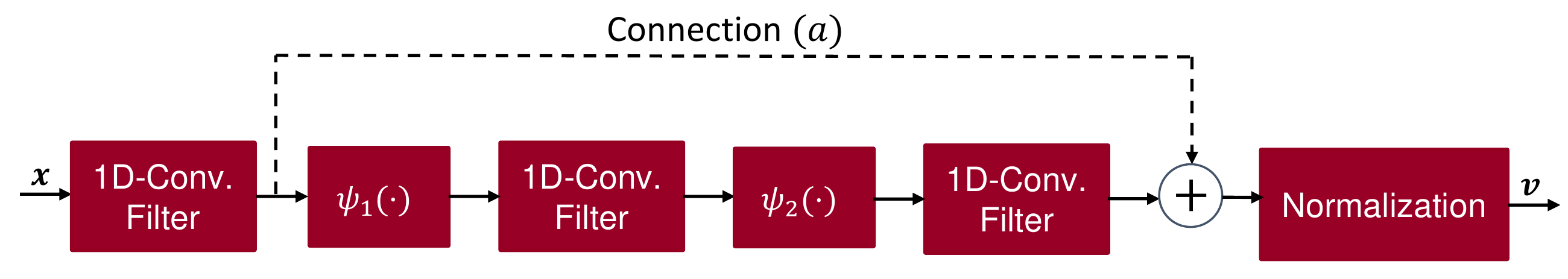}
		\caption{{\small CNN encoder block diagram. The non-linear activation functions $\psi_i$s enable non-linear encoding. The final layer normalizes the output to satisfy the rate and power constraints.}}
		\label{Encoder_block_diagram}
	\end{figure}
	
	\begin{figure}[t]
		\centering
		\includegraphics[width=0.49\textwidth]{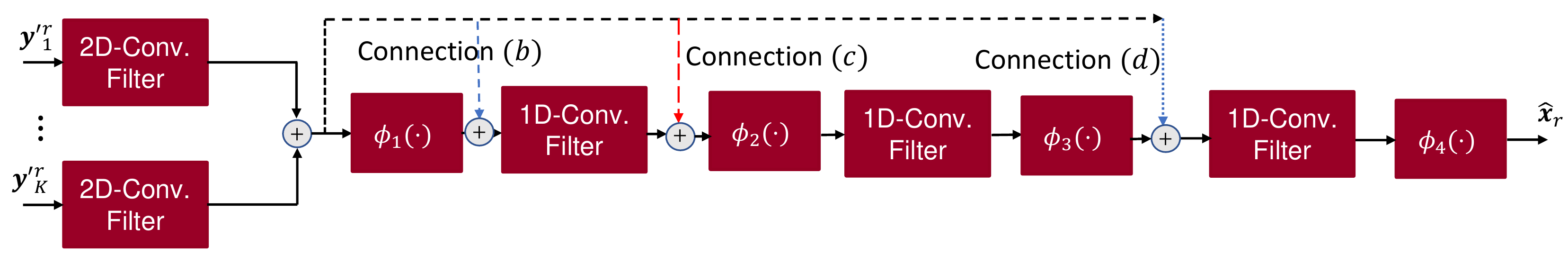}
		\caption{{\small CNN decoder block diagram for user $r$. The first convolutional layer accepts the complex sequences of sufficient statistics for 2D-processing. The outcome sequence is further processed by 1D convolutional layers and non-linear activations.}}
		\label{Decoder_block_diagram}
	\end{figure}
	
	\subsection{Auto-Encoder Architecture}
	
	We propose a CNN-based AE for efficient encoding and decoding for T-NOMA. The CNN AE is shown in Fig. \ref{CNN_AE_NOMA}. At the transmitter, the encoder accepts the input message sequence $\mathbf{x}$ for transmission formed by concatenating the sequences of each user into a vector, i.e., $\mathbf{x} = [\mathbf{x}_1^T,\ldots, \mathbf{x}_K^T]^T$, where $(\cdot)^T$ denotes matrix transpose, and $\mathbf{x}_k$ is an $N \times 1$ sequence intended for user $k$, $k=1,\ldots,K$. The CNN encoder encodes $\mathbf{x}$ into $\mathbf{v}$ for transmission, where $\mathbf{v} = [\mathbf{v}_1^T,\ldots, \mathbf{v}_K^T]^T$, and $\mathbf{v}_k$ is the encoded sequence for user $k$. The encoded sequence $\mathbf{v}$ satisfies rate and power constraints. The encoding rate is one, which is the same rate as the baseline SVD method, i.e., both $\mathbf{x}$ and $\mathbf{v}$ are length-$KN$ sequences. 
	The encoded sequence $\mathbf{v}$ passes through the T-NOMA channel \eqref{Discrete_Time_System_Model_Equation}, with $v_k[n]$ replacing $x_k[n]$.

	At the receiver, each user receives a sequence of sufficient statistics $\mathbf{y}^r = [\mathbf{y}^{rT}_1,\ldots, \mathbf{y}^{rT}_K]^T$, as discussed in subsection \ref{Channel_Model_Subsection}. At user $k$, a CNN decoder estimates the transmitted symbols from the received sequence to give $\hat{\mathbf{x}}_k$. During the training stage, the transmitted sequences and their estimates are used to compute the objective function. The encoder and decoder are jointly trained to minimize the objective function.
	
	\textbf{Encoder Architecture:}
	The encoder block diagram is shown in Fig. \ref{Encoder_block_diagram}. The CNN encoder consists of three finite impulse response (FIR) convolutional filter banks, two non-linear activations, and a normalization stage. The number of filters per filter bank is denoted by $L_{e,i}$, $i=1,\ldots,U_e$, where $U_e$ is the number of convolutional stages at the encoder; $U_e$ = 3 in Fig. \ref{Encoder_block_diagram}. The length of the filters in stage $i$ is denoted by $S_{e,i}$. After each of the first two convolutional stages, a non-linear activation $\psi_i(\cdot)$ is applied to the output. As a result of numerous experiments, we have concluded that the scaled exponential linear unit (SELU) is the most appropriate option for $\psi_i(\cdot)$. SELUs provide self-normalizing properties during training, thereby causing the hidden outputs to have zero mean and unit variance, which ameliorates vanishing or exploding gradients \cite{klambauer2017self}. The SELU activation is defined by the element-wise operation
	\begin{align}
		\text{selu}(x) = \gamma \begin{cases} x, & \text{ if } x>0, \\
			\alpha (e^x - 1), & \text{ if } x\leq 0,
		\end{cases}
	\end{align} 
	where $\gamma \approx 1.0507$ and $\alpha \approx 1.6732$. In addition to the non-linear activation, we also use batch normalization to reduce training performance sensitivity to weight initialization, while enabling some regularization benefits \cite{Batch_Normalization_Sergey}. To ensure the output $\mathbf{v}$ has the expected length of $KN$ estimates, the hidden output samples can be padded with a few zero samples. The final layer is a normalization layer to ensure the transmitted symbols sequence is zero-mean and satisfies the power constraint $P_k = 
 \sum_{n=1}^{N} |v_k[n]|^2/N$. Moreover, we investigate the impact of the linear connection labeled (a) in the Fig. \ref{Encoder_block_diagram}, which provides a more direct path for gradients to flow from the last to the first convolutional layer.

	\textbf{MLP Power allocator:}
	Different from the SVD method, the AE does not necessarily decouple the users' channels. Hence, water-filling
    between channels is suboptimal for the AE.
	 To leverage the end-to-end optimization of the AE for power allocation, we propose MLP-PA that is jointly trained with the encoder and decoder. Such MLP-PA accepts estimates of each user's CSI at the transmitter $\hat{h}_{r,\text{Tx}}$ to allocate the available power to each user's sequence, as shown in Fig. \ref{CNN_AE_NOMA}. The MLP-PA input layer is a size $2K \times M_{h_1}$ matrix that accepts the real and imaginary parts of the $K$ CSI terms and outputs $M_{h_1}$ hidden outputs, which are then passed through the selu($x$) activation. We use two hidden layers for the MLP-PA. The first hidden layer contains a matrix of size $M_{h_1}\times M_{h_2}$ followed by selu($x$) activation. Batch normalization is used after the selu($x$) activation. The second hidden layer contains a matrix of size $M_{h_2}\times M_{h_3}$. The final layer linearly maps the $M_{h_3}$ outputs to $K$ outputs, denoted $o_k$. The matrix coefficients of the input and hidden layers are trainable parameters that are optimized in the end-to-end training process. The soft-max activation is applied to the $K$ hidden outputs to give the ratios of the total powers $P$ that will be allocated to the users. That is, the power allocated to user $k$ is computed as $
		P_k =Pe^{o_k}/\sum_{j=1}^K e^{o_j}$.
	
	\textbf{MLP CSI Combiner:}
	Given the CSI estimate $\hat{h}_{r,\text{Rx}}$ at each user, the received signal is multiplied by $\hat{h}_{r,\text{Rx}}^*/|\hat{h}_{r,\text{Rx}}|^2$, which maximizes the SNR for linear combining \cite{LinearDiversityCombiningTechniques_Brennan}. Ideally, this multiplication cancels out the phase shift due to the fading channel when the CSI is known at the receiver. This strategy is optimal in a local sense, assuming that there is minimal or no IUI remaining in the received signals. We propose using an MLP-T at each user that is optimized based on the end-to-end performance. MLP-T uses the CSI estimate $\hat{h}_{r,\text{Rx}}$ at each user to provide the combining factor $q_r$ for user $r$ as shown in Fig. \ref{CNN_AE_NOMA}. The MLP-T consists of two hidden layers with selu($x$) activation. The first layer accepts the real and imaginary parts of $\hat{h}_{r,\text{Rx}}$. This first layer contains a matrix of size $2\times M_{c_1}$. The second layer contains a matrix of size $M_{c_1} \times M_{c_2}$. Let $q_r$ denote the MLP-T combining factor at user $r$. Then, the output layer consists of a matrix that maps the $M_{c_2}$ hidden outputs to $[\operatorname{Re}\{q_r\}, \operatorname{Im}\{q_r\}]$.
	
	\textbf{Decoder Architecture:} Fig. \ref{Decoder_block_diagram} shows the CNN decoder's architecture for detecting user $r$'s sequence $\mathbf{x}_r$. The input to the decoder is the sequence $\mathbf{y}^{r}$ scaled by the ratio combining factor $q_r$ to give $\mathbf{y}^{\prime r} \triangleq q_r \mathbf{y}^{r}$. Let the number of filters at each convolutional stage be denoted by $L_{d,i}$, and the length of filters at each stage be denoted by $S_{d,i}$, where $i=1,\ldots,U_d$, and $U_d$ is the number of convolutional stages at the decoder; $U_d = 4$ in Fig. \ref{Decoder_block_diagram}. The first convolutional layer uses 2D FIR filters to process the real and imaginary parts of $\mathbf{y}^{\prime r}_{l}$. The 2D convolution produces 1D-sequences for each 2D filter bank for further processing. The output 1D-sequences per 2D filter bank are then summed to give 1D sequences. Then, a non-linear activation $\phi_1(\cdot)$ is applied followed by 1D-convolutional layers and non-linear activations. The SELU is selected as the non-linear activation for $\phi_1(\cdot)$ and $\phi_2(\cdot)$. For $\phi_3(\cdot)$, the hard swish (h-swish) function is used \cite{HswishPaper_SearchingforMobileNetV3_Howard}. 
	Batch normalization is applied at the output of $\phi_i(\cdot)$, $i=1,2,3$.
	The output of the final convolutional layer is passed through a final activation $\phi_4(\cdot)$ to give the estimate $\hat{\mathbf{x}}_r$ of the length-$N$ symbols sequence intended for user $r$. The activation $\phi_4(\cdot)$ depends on the training criteria as discussed in subsection \ref{TrainingCriteriaSubsection}.
	
	\subsection{Training Criteria}
	\label{TrainingCriteriaSubsection}	

	\textbf{Cross-entropy:} Sequence detection can be formulated as a multi-class multi-label classification problem (cf. \cite{menon2019multilabel}). Let $ \mathds{1}_{x[n]=m}$ be the indicator function such that $ \mathds{1}_{x[n]=m} = 1$ if $x[n]=m$ (and zero otherwise), where $m=0,1,\ldots,M-1$, $M$ is the cardinality of the message set for $x[n]$, and $p_{m,n} = \Pr\{x[n] = m\}$ be the soft estimate $\hat{x}[n]$. Then, the CE loss $\mathcal{H}\{x[n], \hat{x}[n]\} $ between the estimated and correct messages is
	\begin{align}
		\mathcal{H}\{x[n], \hat{x}[n]\} = -\sum_{m=0}^{M-1} \mathds{1}_{x[n]=m}\log(p_{m,n}).
	\end{align}

	Consider the $i$th input sequence $\mathbf{x}^{\langle i\rangle}$ in a training mini-batch, $i = 0,\ldots, L-1$, where $L$ is the number of input sequences in a mini-batch. The average CE loss $J_{\text{CE}}$ is computed by averaging over the sequences in the mini-batch:
	\begin{align}
		J_{\text{CE}} = \dfrac{1}{LKN}  \sum_{i=0}^{L-1}\sum_{n=0}^{KN-1}  \mathcal{H}\{x^{\langle i\rangle}[n], \hat{x}^{\langle i\rangle}[n]\}.
	\end{align}

	If $M=2$, then $\phi_4(\cdot)$ can be the sigmoid activation $\text{sigmoid}(x) = 1/(1+e^{-x})$, where the output of the final convolutional layer has length $N$. In this case, the output of the CNN is $p_{0,n}$, $0\leq n \leq N$. If $M>2$, then $\phi_4(\cdot)$ is the soft-max activation $\sigma(\mathbf{z}_n)_m = e^{z_{n,m}} / \sum_{k=0}^{M-1} e^{z_{n,k}} $, where the last convolutional layer produces $M$ output sequences $\mathbf{z} = [\mathbf{z}_n]$ such that $\mathbf{z}_n = [z_{0,n}, \ldots, z_{M-1, n}]$.
	
		\textbf{MSE:} The MSE can be used for training the AE. For MSE training, $\phi_4(\cdot)$ can be the identity function $I(x) = x$. Then, the output is the estimate of the message in the Euclidean distance sense (as opposed to the probability of correct bit estimation in CE training). The sample average squared error $J_{\text{MSE}}$ over the entire sequence in a mini-batch is computed as follows:
	\begin{align}
		J_{\text{MSE}} = \dfrac{1}{LKN}  \sum_{i=0}^{L-1}\sum_{n=0}^{KN-1}  (x^{\langle i\rangle}[n]-\hat{x}^{\langle i\rangle}[n])^2.
	\end{align}
	In the case where $M = 2$, the tanh activation can also be used for $\phi_4(\cdot)$ instead of $I(x)$.

	\textbf{Cross-entropy with Q-function}
In comparison with the MSE loss, the CE loss has been shown to produce lower BERs. Nevertheless, achieving minimum CE does not guarantee achieving minimum BER.
We use first and second order statistics of the LLRs in each mini-batch to further decrease the BER during training. Assume the transmitted symbols follow the BPSK modulation. The LLR is defined as
\begin{align}
	\text{LLR}_i &=  \log\Big(\dfrac{\Pr \{x_i = +1\}}{\Pr \{x_i = -1\}}\Big) = \log\Big(\dfrac{\hat{p}_i}{1-\hat{p}_i}\Big).
\end{align}

		For each mini-batch, assume the LLRs are Gaussian. Our goal is to minimize the overlap between the Gaussian corresponding to a $+1$ transmitted bit and the Gaussian corresponding to a $-1$ transmitted bit. This overlap is captured by the Q-function, where $Q(x) \triangleq (1/\sqrt{2\pi}) \int_{x}^{\infty}e^{-u^2/2}du$. 

In a mini-batch containing $L_s$ sequences of length-$N_s$ bits per user, we compute means and variances of the conditional LLRs. Let $s\in \{-1,+1\}$ denote the transmitted symbol. Each mini-batch contains $L_sN_s$ LLRs corresponding to symbol $s$, where $L_{+1} + L_{-1} = L_s$, and $N_{+1} + N_{-1} = N_s$.
Let LLR$^s_{a,b}$ denote the conditional LLRs corresponding to symbol $s$, where $a \in \{0,\ldots,L_s-1\}$, and $b \in \{0,\ldots,N_s-1\}$. 
Then, the empirical mean and variance of LLR$^s_{a,b}$ are computed as	
	\begin{align}
	\mu_s = \dfrac{1}{L_sN_s}\sum_{a,b} \text{LLR}^s_{a,b},\text{ and } \sigma^2_s = \dfrac{1}{L_sN_s}\sum_{a,b} (\text{LLR}^s_{a,b}- \mu_s)^2. \label{mean_LLRs_Eq}
	\end{align}
We introduce the training Q-function loss (Q-loss) as: 
\begin{align}
		J_{\text{Q}} \triangleq \dfrac{1}{2}\sum_{s} Q\bigg(\kappa\dfrac{\mu_s}{\sigma_s}\bigg), \label{Q_loss_equation}
\end{align}
where $\kappa > 0$ is a hyperparameter to be tuned. 
The combined cross-entropy with Q-loss objective function is given by:
	\begin{align}
		J_{\text{CEQ}} = \mathcal{H}\{x[n], \hat{x}[n]\} + \alpha J_{\text{Q}},
    \label{Q_loss_objective}
	\end{align}
	where $\alpha$ is a weight factor. Appendix~\ref{App A} provides a derivation of the gradient of $J_Q$ with respect to the final layer AE weights, which can be used in AE training. Empirically, we find that small values of $\alpha$ in the range $(0,1)$ can provide significant BER improvements for appropriate values of $\kappa$. The Q-loss in \eqref{Q_loss_equation} can be extended to $M$-QAM and $M$-PSK modulations by defining  $M$-ary LLRs (cf. \cite{LLR_QAM_Park_2015}).
	 
	\begin{table}[t]
		\centering
		\caption{\small Number of convolutional filters $L_{e,i}$ and $L_{d,j}$ for the encoders and decoders of different variants of the CNN AE, $i=1,2,3$ and $j=1,2,3,4$. All CNN AEs user FIR filters of length 11 samples, i.e., $S_{e,i} = S_{d,j} = 11$.}
		\label{CNN_Variant_Parameters}
		\setlength{\tabcolsep}{7pt}
		\begin{tabular}{@{}ccc@{}}
			\toprule
			Method& Encoder & Decoder\\ \midrule
			CNN AE1 & $1,1,1$           & $1,1,1,1$        \\
			CNN AE2 & $2,4,2$             & $1,1,1,1$        \\
			CNN AE3 & $2,4,2$		      & $1,1,2,2$          \\
			CNN AE4 & $1,1,1$		       & $2,4,4,2$         \\ 
			CNN AE5 & $2,4,2$		       & $2,4,4,2$         \\ 
			CNN AE6 & $1,1,1$		       & $1,2,8,4$         \\
			CNN AE7 & $1,1,1$		       & $2,16,128,16$         \\
			CNN AE8 & $2,4,2$		       & $2,16,128,16$         \\  
			CNN AE9 & $2,4,2$		       & $4,32,128,16$    \\       
			\bottomrule
		\end{tabular}
	\end{table}

	\subsection{Complexity Comparison}
	\label{Complexity_comparison_section}
		\begin{table*}[t]
		\centering
		\caption{\small Complexity comparison for sequence estimation. The numerical values give the complexity order for the simulated system with $K=2$ and $N=512$, and CNN AE5. The CNN AE also requires $O(KN)$ and $O(K^3N)$ computations of the non-linear activations for the encoder and decoder, respectively. For MLP-PA we use $M_{h_i} = 32$. For MLP-T, we use $M_{c_i} = 8$.}
		\label{ComplexityTable}
		\setlength{\tabcolsep}{2pt}
		\begin{tabular}{@{}ccc@{}}
			\toprule
			Method& FLOPs & Storage\\ \midrule
			SVD, Encoder & $(KN)^2$ $\sim 1.0486\times 10^6$          & $(KN)^2$ $\sim 1.0486\times 10^6$       \\
			SVD, Decoder & $(KN)^2 $ $\sim 1.0486\times 10^6$            & $(KN)^2$ $\sim 1.0486\times 10^6$       \\
			CNN AE, Encoder      & $KN\sum_{i=1}^{U_e} L_{e,i}S_{e,i}$	$\sim 1.8022\times 10^5$	            & $K\sum_{i=1}^{U_e} L_{e,i}S_{e,i}$ $\sim 352$          \\
			CNN AE, Decoder      & $K^2N(2KL_{d,1}S_{d,1} + \sum_{i=2}^{U_d} L_{d,i}S_{d,i})$ $\sim 4.0550\times 10^5$		             & $K(2KL_{d,1}S_{d,1} + \sum_{i=2}^{U_d} L_{d,i}S_{d,i})$ $\sim 396$  \\
			MLP-PA &$M_{h_1}(2K + M_{h_2}) + M_{h_2}M_{h_3}+M_{h_3} \sim 656$ &$M_{h_1}(2K + M_{h_2}) + M_{h_2}M_{h_3}+M_{h_3} \sim 656$\\
			MLP-T &$2M_{c_1} +M_{c_1}M_{c_2} +2M_{c_2} \sim 96 $ &$2M_{c_1} +M_{c_1}M_{c_2} +2M_{c_2} \sim 96$
			\\ \bottomrule
		\end{tabular}
	\end{table*}
	Table \ref{ComplexityTable} summarizes the complexity and storage requirements for the SVD-based baseline and the CNN AE employed in the T-NOMA system. The baseline pre-multiplies the $KN\times 1$ vector $\mathbf{Px}$ by a $KN \times KN$ matrix and post-multiples the received signal sequence $\mathbf{y}^r$ by another $KN \times KN$ matrix. Hence, the order of the number of floating-point operations (FLOPs), i.e., elementary multipliers and adders, for each matrix multiplication is $O((KN)^2)$ per sequence encoding and decoding. Whereas, the CNN AE uses discrete-time convolutions, which require $O(KNL_{e,i}S_{e,i})$ and $O(K^2N{L_{d,i}S_{d,i}})$ FLOPs for each convolutional layer in the encoder and the decoder, respectively. Therefore, the CNN AE's complexity is linear in $N$, whereas the SVD's complexity is quadratic in $N$.
	
	The SVD method stores $(KN)^2$ parameters at the encoder and decoders. The CNN AE only stores the FIR filter coefficients at each convolutional stage and the batch normalization means and variances, giving a storage order of $KL_{e,i}S_{e,i}$ and $K^2L_{d,1}S_{d,1}$ for the encoder and decoder, respectively. For the CNN decoder, where $K$ is large, the complexity is dictated by the first convolutional layer because it processes the length-$KN$ complex sequence of sufficient statistics. Thus, compared with the SVD's quadratic storage order in $KN$, the CNN AE's storage is independent of $N$ and quadratic in $K$. 	

 \section{Numerical Results}
	\label{NumericalResultsSection}
  \begin{figure}[t]
		\centering
	\includegraphics[width=8.7 cm, height=6.2cm]{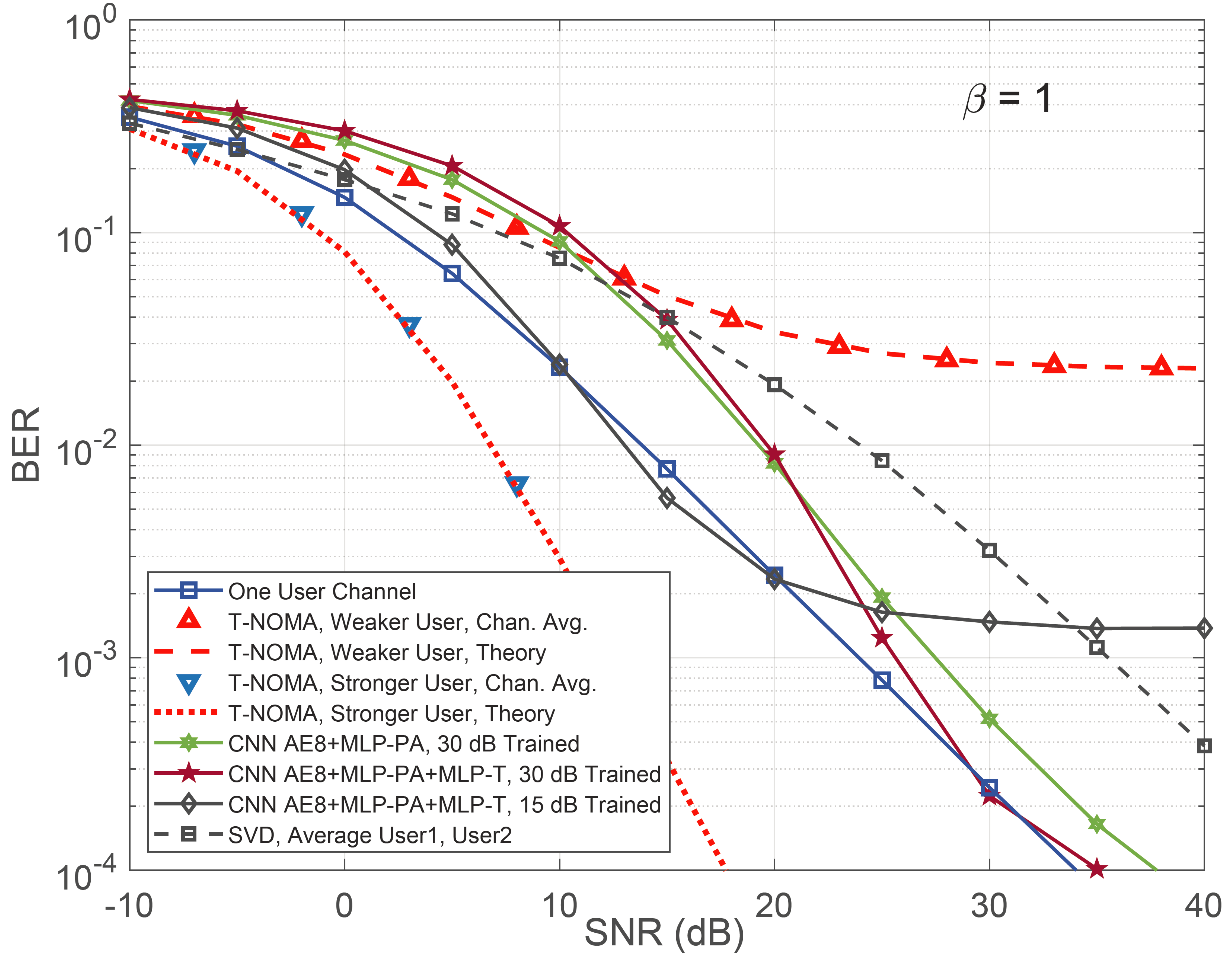}
		\caption{{\small BER comparison between SVD T-NOMA system simulation, one-user Rayleigh fading channel, T-NOMA with stronger/weaker user selection (per Appendix \ref{AverageBERAnalysisAppendix}), and CNN AE T-NOMA system simulation. All simulations assume perfect CSI and timing. Given the same average SNR per user, the CNN AE8 communicates at double the rate of a single-user system while maintaining the same BER when trained at SNRs of 15 and 30 dB.}}
  \label{BER_TNOMA_and_PNOMA}
	\end{figure}

We present simulation results comparing the BER performances of the discussed systems. We consider a two-user NOMA system, i.e., $K=2$, with BPSK modulation for both users. The fading is Rayleigh distributed such that $\mathbf{E}\{|h_i|^2\} = 1$, $i=1,2$. Both users have the same variance of the AWGN electronics noise, i.e., $\sigma_1^2 = \sigma_2^2 = N_0$. The SNR is computed with respect to user 1's signal, i.e., SNR $ = \mathbf{E}\{ |h_1|^2 / \sigma_1^2\} = 1/N_0$. The frame length $N = 512$ data symbols per user. Testing is performed using $64\times 2048 = 131,072$ frames. We use the root raised cosine pulse (RRCP) with a unity roll-off factor for $p(t)$. The discrete time RCP samples $\mathbf{g}_{l,k}$ are sampled accordingly and truncated to have an ISI span of 7 symbols per user, which captures most of the energy in the RCP. For T-NOMA, we use the timing offset $\tau_{\text{design}} = T/2$, which is optimal for two-user NOMA \cite{ExactBERPerformanceforSymbolAsychronousTwoUserNOMA_Liu}.

Average BER results with BPSK modulation for T-NOMA with stronger/weaker user selection are also provided for comparison; results are obtained by numerically averaging $\text{BER}_i=Q(\sqrt{2\eta_i})$ for several random realizations of the instantaneous channel SNRs $\eta_i$, where $i\in\{s,w\}$ corresponds to the stronger and weaker user, respectively. We use $\eta_s=P_s|h_s|^2/N_0$ and $\eta_w=(P_w|h_w|^2)/(P_sG_{2,1}|h_w|^2+N_0)$, where the channel gains have been sorted such that $|h_s|^2/\sigma_s^2>|h_w|^2/\sigma_w^2$. The theoretical BER curves for T-NOMA with stronger/weaker user selection are obtained by using \eqref{EQ_B4} and \eqref{EQ_B5} in Appendix \ref{AverageBERAnalysisAppendix}. For all other T-NOMA results, we simulate a T-NOMA system per subsection \ref{Channel_Model_Subsection}. For the SVD method, power allocation is performed using the water-filling algorithm described in subsection \ref{AchievableRatesSection}. We use at least 1024 frames for testing.

\begin{table}[t]
	\centering
	\caption{\small Hyperparameter Settings.}
	\label{Hyperparameters_Settings_Table}
	\begin{tabular}{@{}ll@{}}
		\toprule
		Hyperparameter                               & Setting                   \\ \midrule
		Number of Data Symbols per Sequence per User & $512$                       \\
		Number of Training Sequences                 & $131,072$                   \\
		Number of Validation Sequences               & $512$                       \\
		Number of Testing Sequences                  & $131, 072$                  \\
		Number of Sequences per Training Minibatch      & $32$                        \\
		Learning Rate                                & $3\times 10^{-3}$           \\
		Number of Training Epochs                    & $20$                        \\
		Training Algorithm                           & Adam \\
		Training SNR								 &$30$ dB \\
		Design Time Offset                           & $T/2$                     \\
		Number of Users                              & $2$                         \\
		Fading Distribution                          & Rayleigh                  \\
		Distribution of Timing Error                 & Uniform                   \\
		Distribution of CSI Error                    & Gaussian                  \\ \bottomrule
	\end{tabular}
\end{table}
\begin{figure}[t]
	\centering
\includegraphics[width=8.250 cm, height=5.0cm]{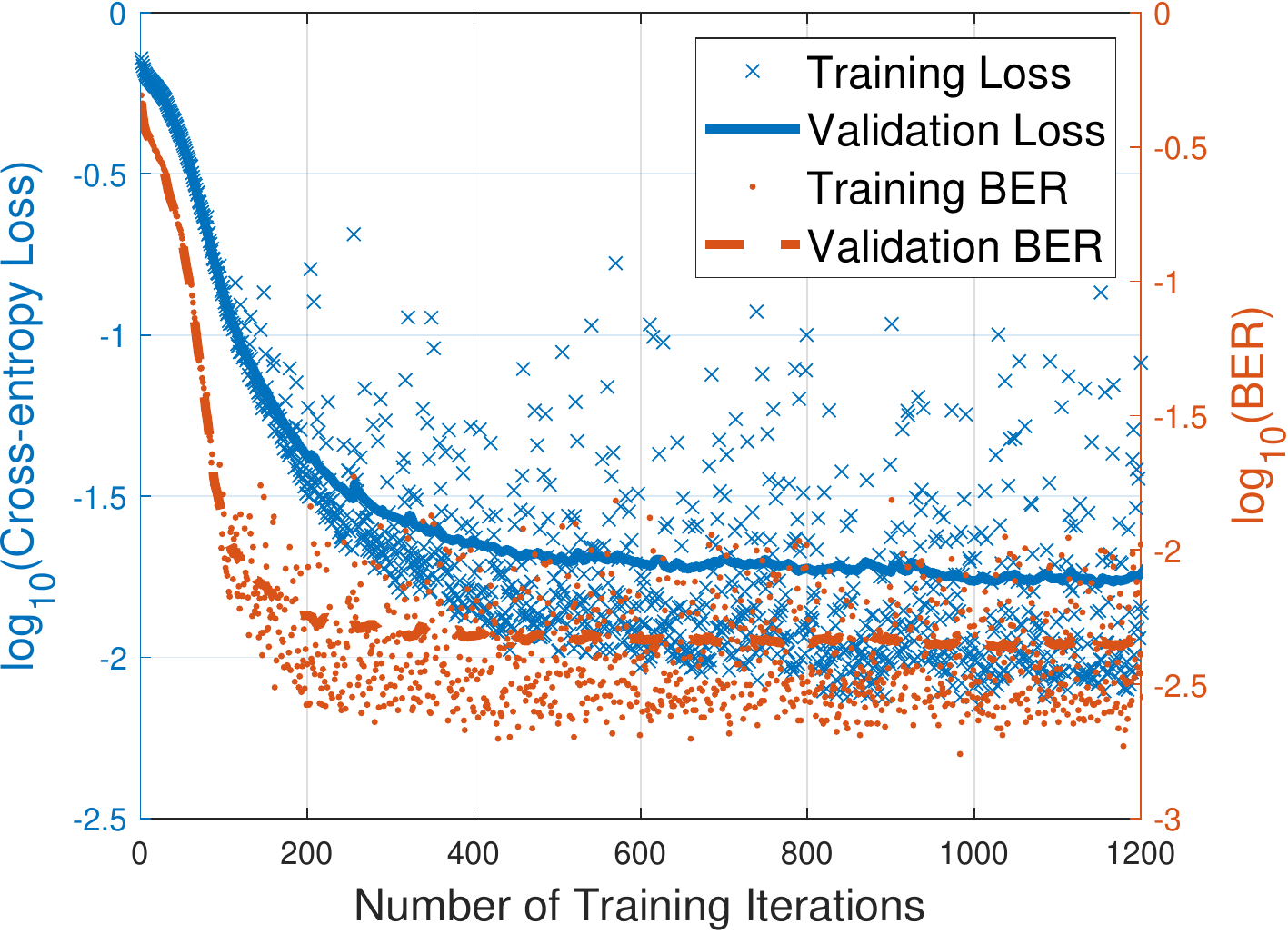}
	\caption{{\small Learning curve.}}
	\label{LearningCurveFig}
\end{figure}
%
\subsection{BER Performance in T-NOMA with SVD, T-NOMA with CNN AE, and T-NOMA with user selection}
This subsection investigates the BER performance of T-NOMA with SVD, with CNN AE, and T-NOMA with user selection, under perfect CSI and no timing error. 

Fig.~\ref{BER_TNOMA_and_PNOMA} provides a BER performance comparison of the considered T-NOMA systems.
In T-NOMA with user selection, the stronger user attains the best BER results, but the weaker user's performance shows an error floor, plateauing at approximately 0.02 BER by 20 dB. By contrast, the SVD in T-NOMA enables equal average BER performance for both users, ensuring fairness. 

Figure~\ref{BER_TNOMA_and_PNOMA} also depicts the BER performance of a single-user system utilizing ML detection in a Rayleigh fading channel. At an SNR of 30 dB, we observe a 9 dB gain by the auto-encoder architecture AE8, employing an MLP-PA, over the SVD method. Integrating the MLP-T with the CNN AE8 yields an additional 3 dB performance improvement. The CNN AE system, when trained at 15 dB, achieves substantial gains over the SVD across SNRs from 5 dB to 30 dB. Given the same average SNR per user, the CNN AE8 achieves twice the communication rate of a single-user system while preserving the same BER, particularly when trained at SNRs of 15 dB and 30 dB. Similarly, the SVD baseline achieves double the rate of a single user, although at a marginally higher BER.

\begin{figure}[t]
	\centering\includegraphics[width=8.70 cm, height=6.2cm]{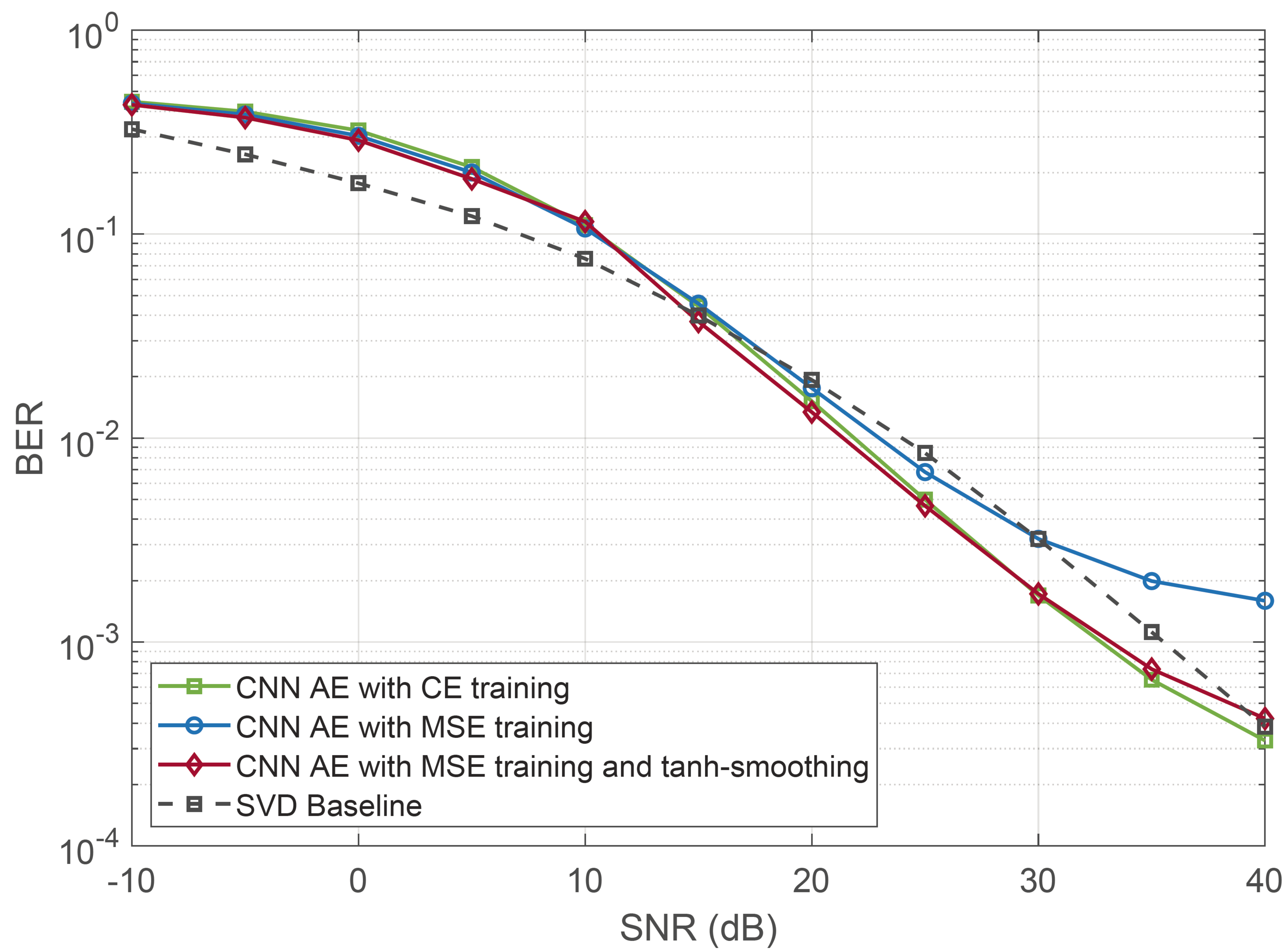}
	\caption{{\small BER performance comparison between the SVD baseline and the CNN AE with different training modes.}}
	\label{BER_Comparison_Figure}
\end{figure}

\subsection{SVD baseline vs. CNN AE for T-NOMA}
\label{CNNvsSVD_Results_Subsection}

\begin{figure}[t]
	\centering
	\includegraphics[width=8.70 cm, height=6.2cm]{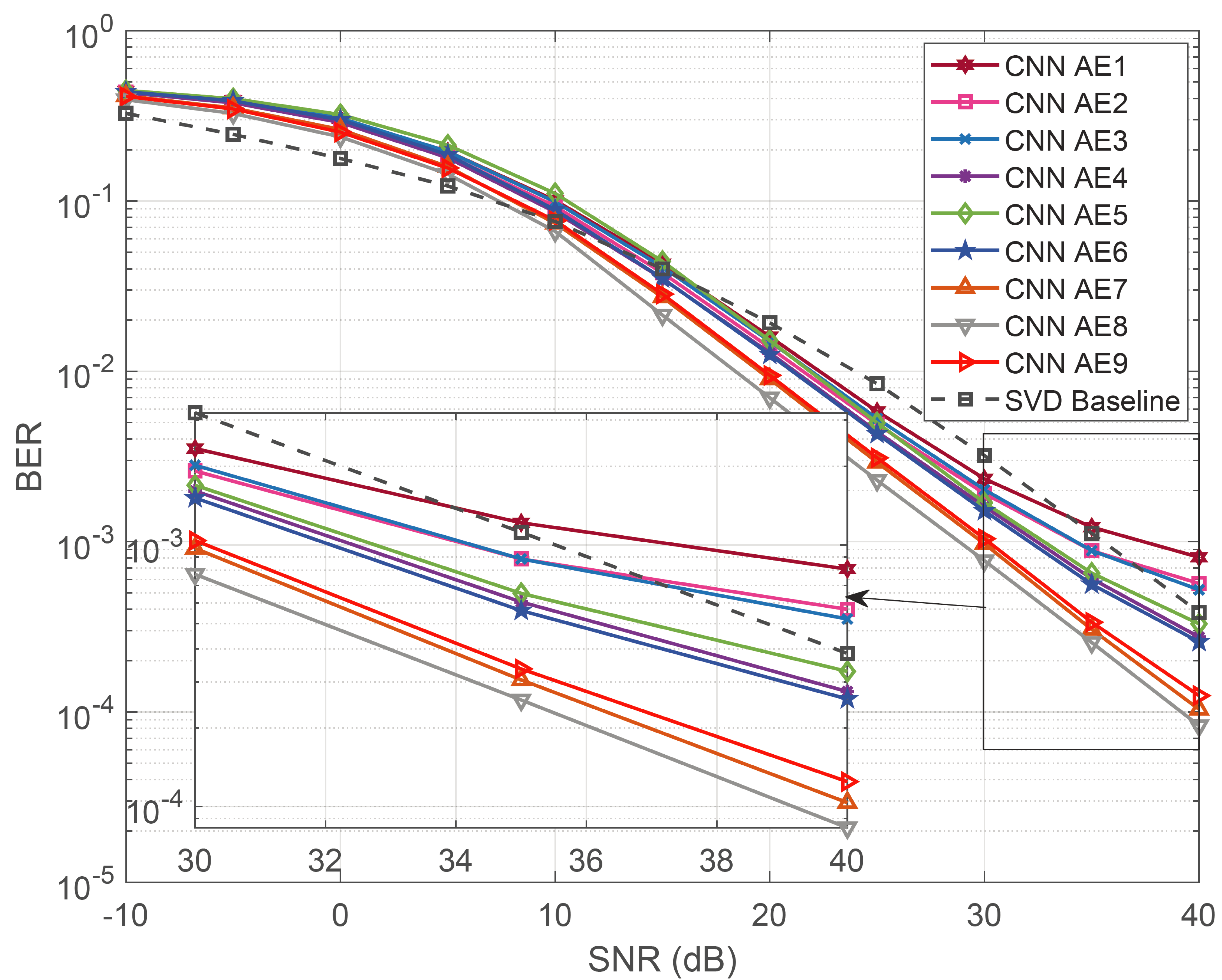}
	\caption{{\small BER performances for several CNN AE variants. The complexity increases from CNN AE1 to CNN AE9.}}
	\label{Different_Architectures_BER_Comparison_Figure}
\end{figure}
We compare the T-NOMA BER performances of the SVD and the CNN AE under various training modes. The CNN AE is trained using randomly generated $ 131,072$ frames at 30 dB SNR and tested using another random set of $131,072$ frames per testing SNR; each frame has 512 bits per user. The Adam optimizer, a variant of the stochastic gradient descent algorithm, is used in the back propagation \cite{Adam_Kingma}. We train the CNN AE for 20 epochs using a fixed base learning rate of $3\times 10^{-3}$. We implemented the CNN AE system in Python using the Pytorch library. Table \ref{Hyperparameters_Settings_Table} summarizes the simulation settings.  
Fig \ref{LearningCurveFig} shows the progressions of the BER and the CE loss with training iterations. We observe that minimizing the CE also reduces the BER.

Initially, we consider the scenario without timing or with imperfect CSI estimation. The CNN AE includes the encoder and decoder CNNs, excluding the MLP components unless specified. The default power allocation for the CNN AE is $P_1 = P_2 = 1$ unless otherwise stated. Fig.~\ref{BER_Comparison_Figure} compares BER performance with three different training modes. CE-based training of the CNN AE yields better BER than training on the MSE loss. However, utilizing a tanh activation function improves the BER when MSE is used.

 \begin{figure}[t]
		\centering
	\includegraphics[width=8.7 cm, height=6.1cm]{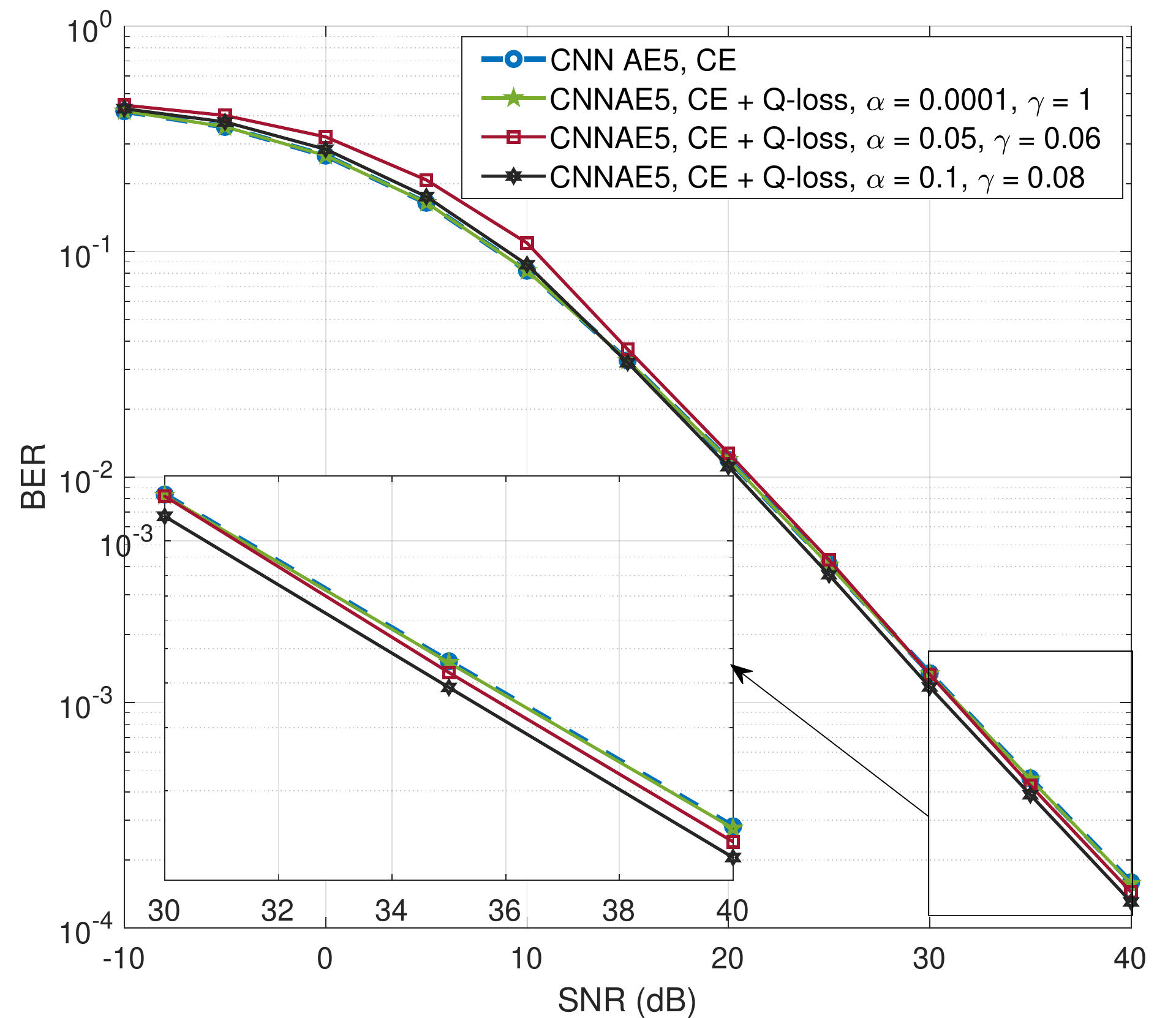}
		\caption{{\small Impact of Q-function loss.}}	\label{CNNAE5_Comparison_withQ_loss}
\end{figure}

\subsection{BER Performance of Different CNN AE Variants}
This subsection illustrates the CNN AE's performance-complexity trade-off when no timing or imperfect CSI estimation are present. Fig.~\ref{Different_Architectures_BER_Comparison_Figure} shows the BER performances of the different CNN AE variants shown in Table \ref{CNN_Variant_Parameters}. CNN AE1 uses the fewest number of filters and requires the lowest complexity. CNN AE9 entails the highest computational complexity among the variants.
Interestingly, CNN AE4 uses fewer filters at the encoder than CNN AE5 and achieves similar BER performance. This suggests that increasing the decoder's complexity can yield higher performance improvements than increasing the encoder's complexity. The best performance is achieved by CNN AE8, which outperforms the SVD by about 8 dB at an SNR of 30 dB. We note that with further tuning of the hyperparameters using Bayesian optimization (cf. \cite{MakingaScienceofModelSearchHyperparameterOptimization_Bergstra}), CNN AE9 may perform at least as well as CNN AE8.

\begin{figure}[t]
\centering\includegraphics[width=8.7 cm, height=6.2cm]{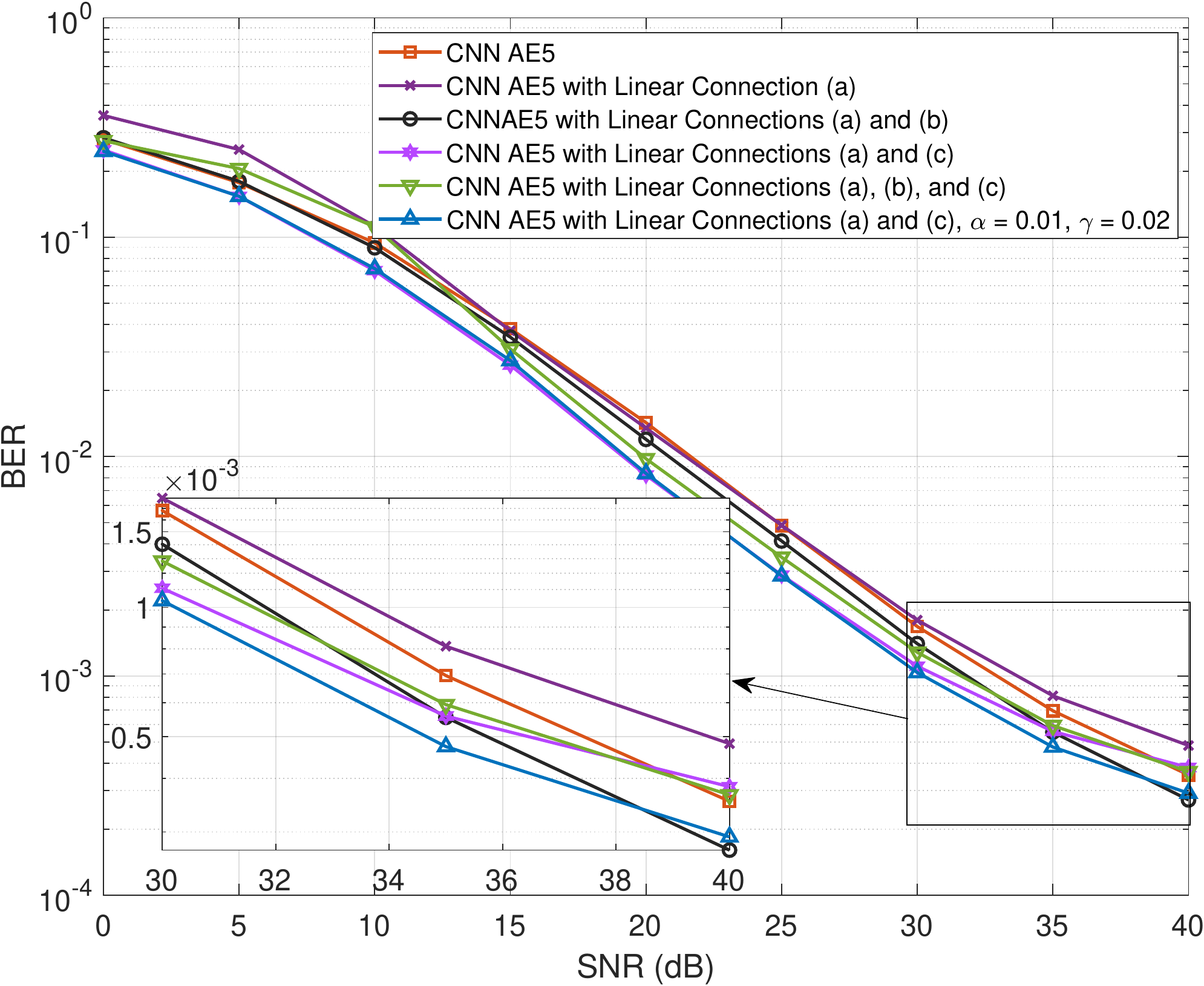}
\caption{{\small Impact of linear connections.}}
\label{CNNArchitectureswithLinearConnection}
\end{figure}
%
\subsection{Benefits of Q-function Loss}
Fig. \ref{CNNAE5_Comparison_withQ_loss} shows CNN AE5's BER performance with different loss functions. By combining the CE with the Q-function loss, we achieve around 0.8 dB gain for a BER of $2\times 10^{-4}$. To achieve this gain, the hyperparameters $\kappa$ and $\alpha$ in \eqref{Q_loss_equation} and \eqref{Q_loss_objective} must be tuned appropriately. This BER gain can be achieved at no increase in inference complexity, at the cost of using the combined CE and Q-function loss during training.

\begin{figure}[t]
\centering
\includegraphics[width=8.8 cm, height=6.2cm]{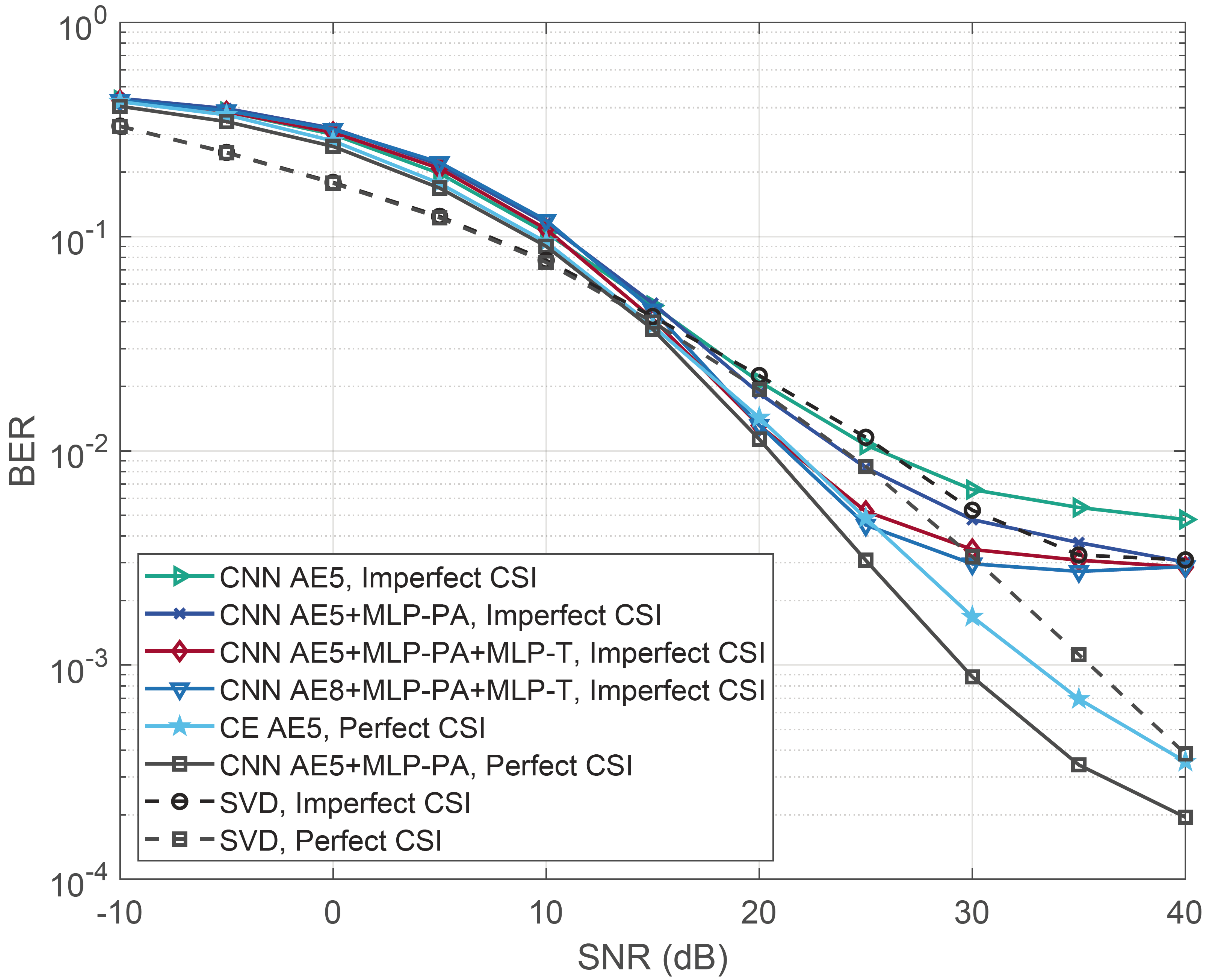}
	\caption{{\small Impact of imperfect CSI on BER performance. The error on the CSI is added at 20 dB, i.e., $\mathbf{E}[|\delta|^2] = 0.01$.}}	\label{ImpactofImperfectCSIFigure}
	\end{figure}
\begin{figure}[t]
\centering
\includegraphics[width=8.7 cm, height=6.3cm]{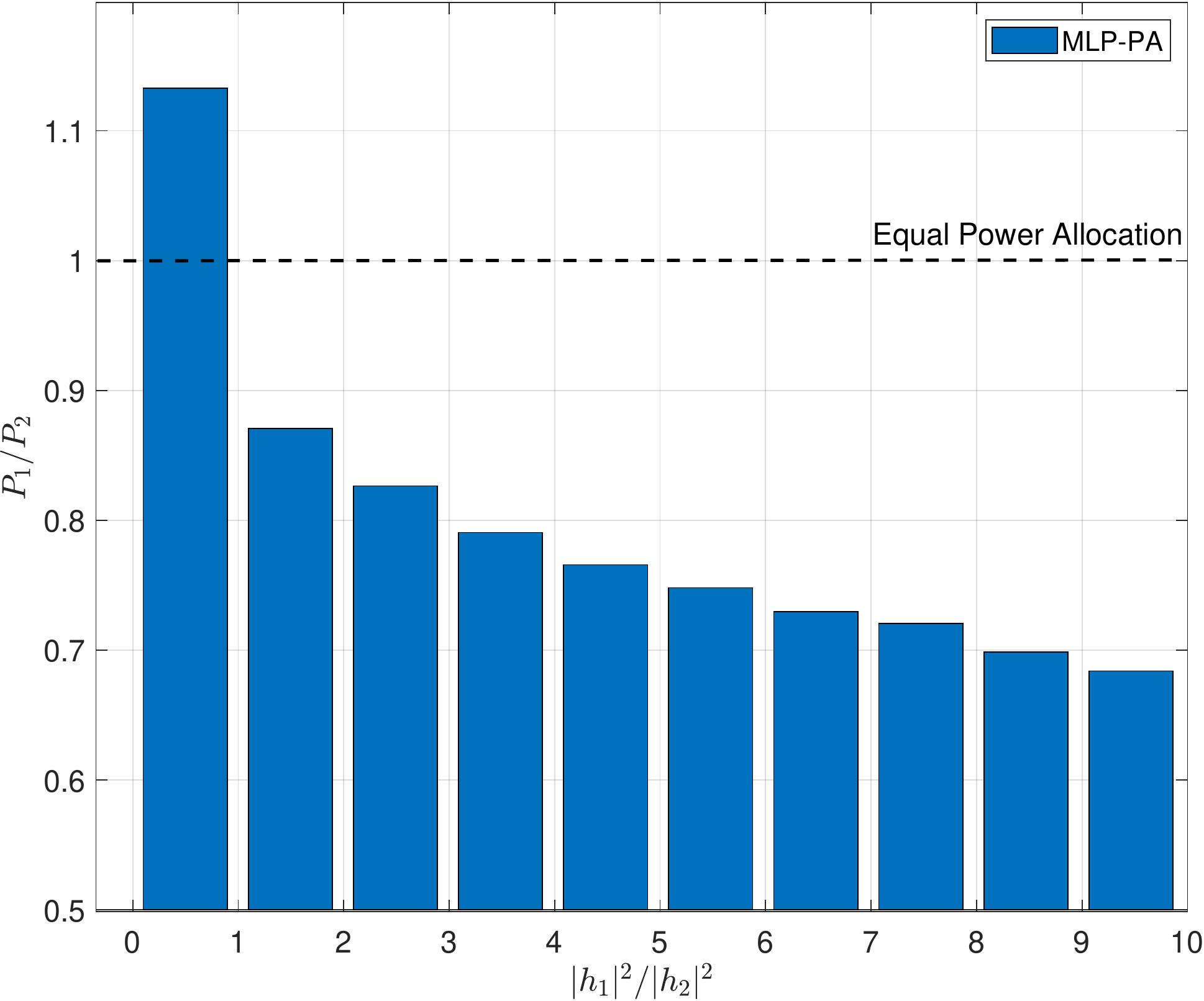}
	\caption{{\small Power allocation using MLP power allocator for different realizations of the CSI.}}
\label{MLP_Power_Allocation_Figure}
\end{figure}
\subsection{Benefits of Linear Connections}
%
Fig. \ref{CNNArchitectureswithLinearConnection} illustrates CNN AE5's BER performance with different linear connection combinations at encoder and decoder. CNN AE5 with linear connections (a) and (c) in Fig.~\ref{Decoder_block_diagram} provides around 2.2 dB gain for a BER of $1.1\times 10^{-3}$ compared with the original CNN AE5, and also achieves the lowest BER at 30 dB. When the combined CE and Q-function loss is used for training, the gain over the best performing system trained only with CE is between 0.5 dB and 1 dB for SNRs between 30 dB and 38 dB.
\subsection{Impact of Imperfect CSI}	
 	\begin{figure}[t]
		\centering
	\includegraphics[width=8.5 cm, height=6.0cm]{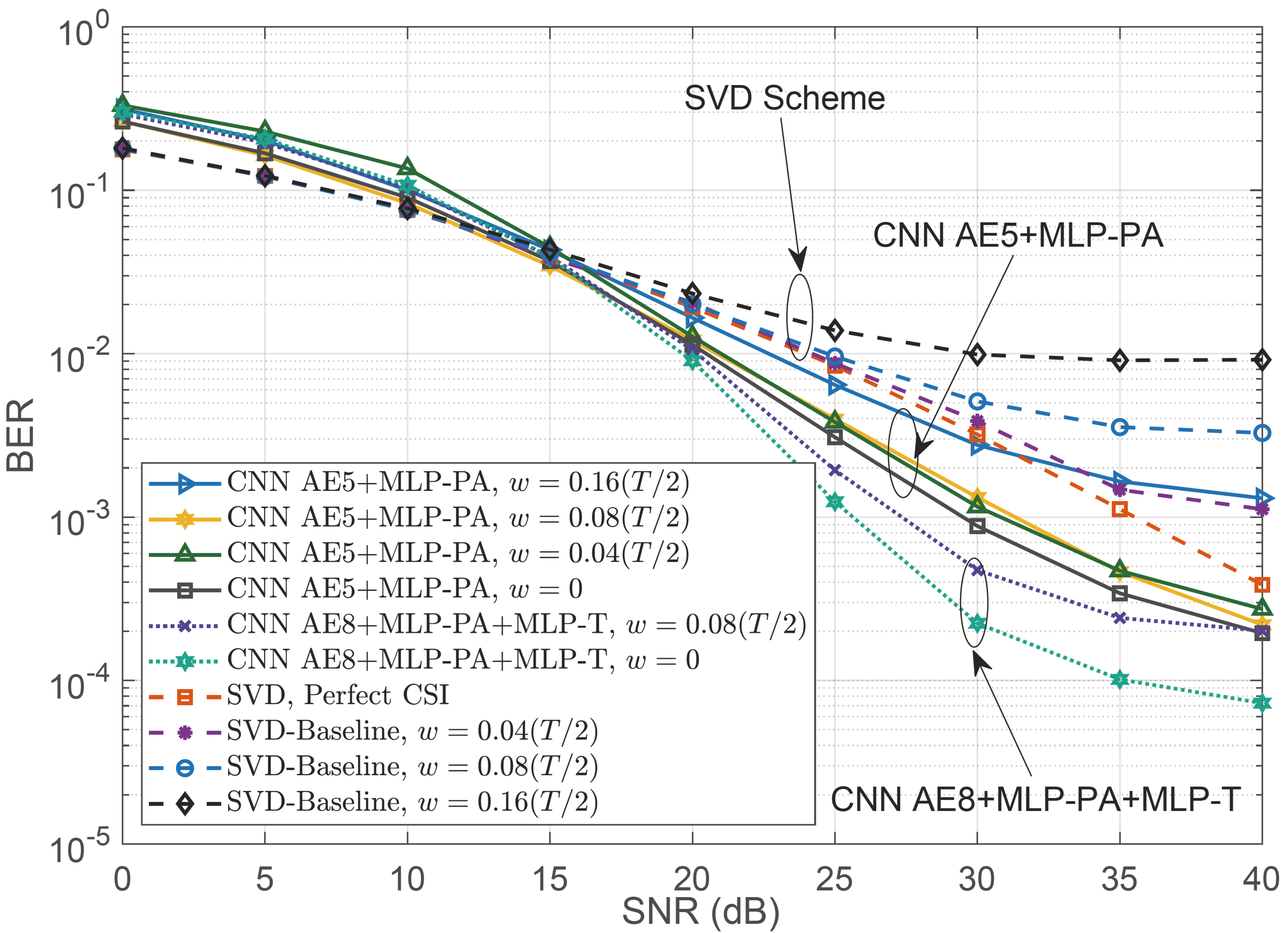}
		\caption{{\small Impact of imperfect timing offset error on the BER performance. The uniform timing error $\epsilon$ is added at width of $w\times$ the design offset $\tau_{\text{design}}$, i.e., $\epsilon\sim U(-(w/2)\tau_{\text{design}},(w/2)\tau_{\text{design}})$. }}
		\label{ImpactofTimingOffsetErrorFigure}
	\end{figure}

This subsection investigates the impact of imperfect CSI at the transceivers on the BER of the SVD and the CNN AE when no timing errors are present.

Fig.~\ref{ImpactofImperfectCSIFigure} shows the impact of imperfect CSI on the SVD and CNN AE performances. Imperfect CSI causes error floors in the BERs of all methods. The error in the CSI estimate causes data dependent noise and also affects the AWGN, as shown in \eqref{ImpactTEandImperfectCSIEquation}. The MLP-PA improves the CNN AE's resilience to imperfect CSI. Including the MLP-T allows further performance gains by the CNN system. For SNRs between 20 dB and 30 dB, CNN AE5 with MLP-PA and MLP-T provides up to 7 dB performance gain over the SVD baseline.	
Fig.~\ref{MLP_Power_Allocation_Figure} illustrates a histogram of the power allocation between users by the MLP-PA. 
\subsection{Impact of Timing Offset Error}
%
 Fig.~\ref{ImpactofTimingOffsetErrorFigure} shows the impact of random timing errors on the BER of the SVD baseline and CNN AE5 with MLP-PA, assuming perfect CSI. Timing errors cause about 5 dB loss in performance for the SVD. By comparison, CNN AE5 experiences about a 2 dB loss in high-SNR performance. CNN AE5 outperforms the SVD baseline by significant margins at high SNRs when timing error is introduced, e.g., by about 10 dB at a BER of $2\times 10^{-3}$. CNN AE8 gives an additional 3 dB gain over CNN AE5 at that same BER. Hence, the CNN AE is more robust to timing errors than the SVD method.

\subsection{Impact of Timing Error and Imperfect CSI}
\begin{figure}[t]
	\centering	\includegraphics[width=8.5 cm, height=6.0cm]{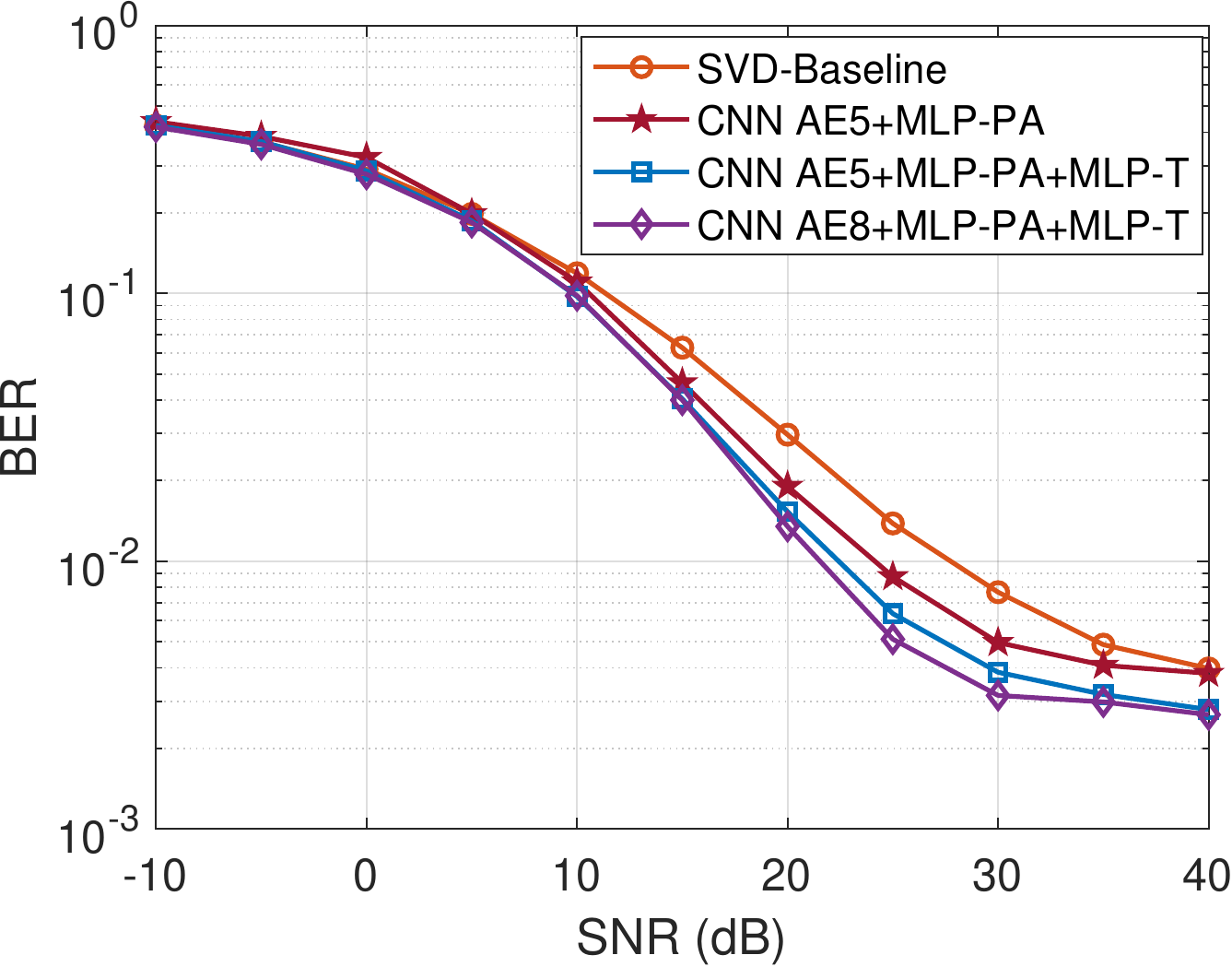}
	\caption{{\small BER performance comparison between SVD and CNN AE where the CSI error variance is 0.01, and the timing error width is at $16\%$ of $\tau_{\text{design}}$. }}
\label{ImpactTEandImperfectCSI_Sampling_Figure}
\end{figure}
\begin{figure}[t]
	\centering
\includegraphics[width=8.5 cm, height=5.90cm]{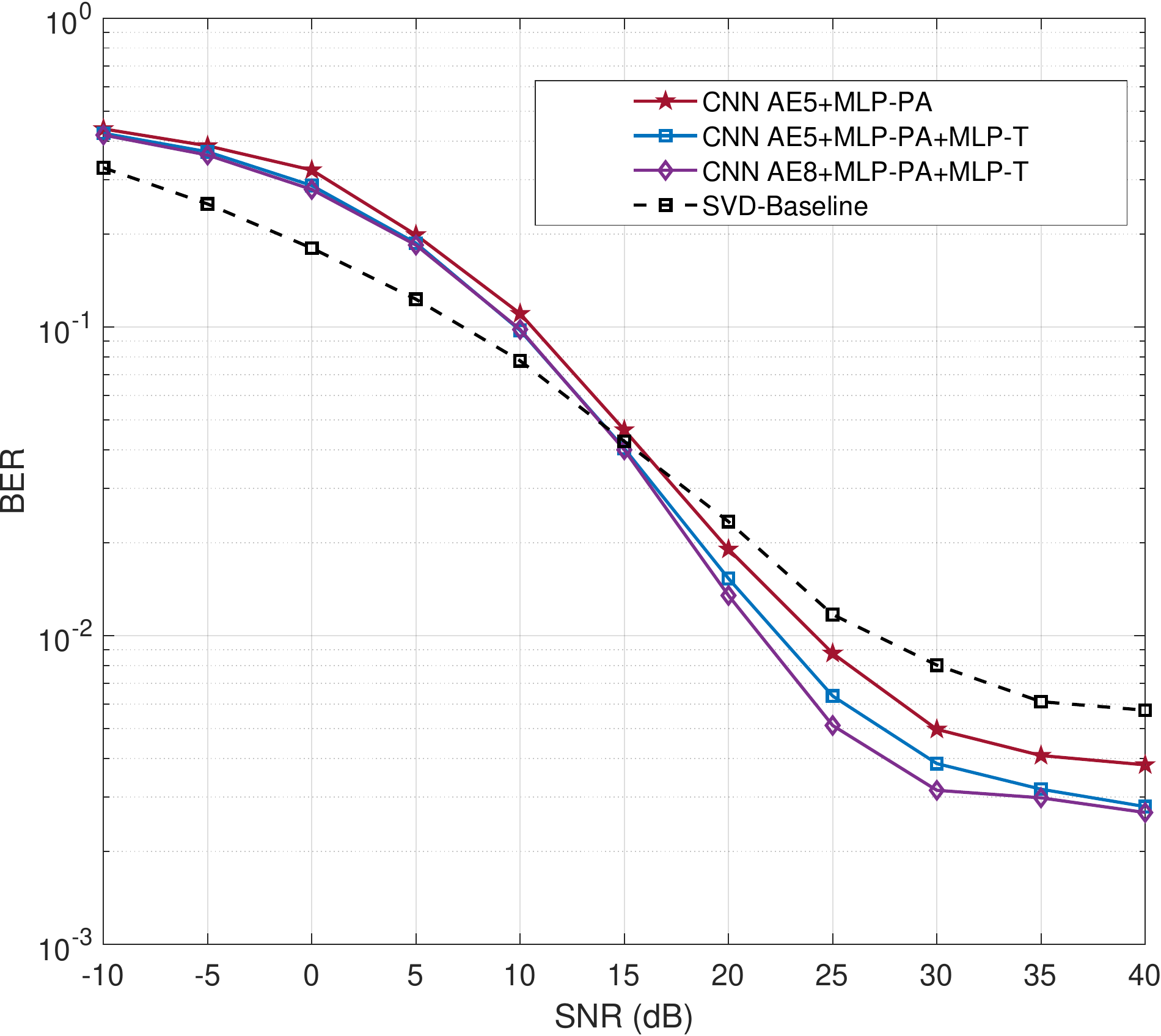}
	\caption{{\small BER comparison, impact of CSI error and timing error. }}
\label{BER_comparison_impact_of_CSI_error_and_timing_error}
\end{figure}

 Fig.~\ref{ImpactTEandImperfectCSI_Sampling_Figure} compares the BERs of CNN AE5 and the SVD method when both timing error and imperfect CSI are considered. CNN AE5 with the MLP-PA achieves about 2 dB performance gain over the SVD for SNRs between 20 dB and 30 dB. Integrating the MLP-T with CNN AE5 achieves an additional 1 dB to 2 dB gains for SNRs between 20 dB and 40 dB. CNN AE8 achieves an additional 2 dB gain at 25 dB SNR. Both CNN AEs achieve approximately the same BER by an SNR of 40 dB. Also, at BERs of $4\times10^{-3}$ and $3\times10^{-3}$, CNN AE5 and CNN AE8 with MLP-PA and MLP-T achieve gains of 10 dB and greater than 10 dB over SVD, respectively. The SVD power allocation is optimized for each SNR, whereas the AE was trained on 30 dB, resulting in better performance of the SNRs less than 10 dB. Additionally, such under performance at low SNRs suggests that the AE can benefit from optimizations for low SNRs. 
\begin{figure}[t]
	\centering
\includegraphics[width=8.7 cm, height=5.2cm]{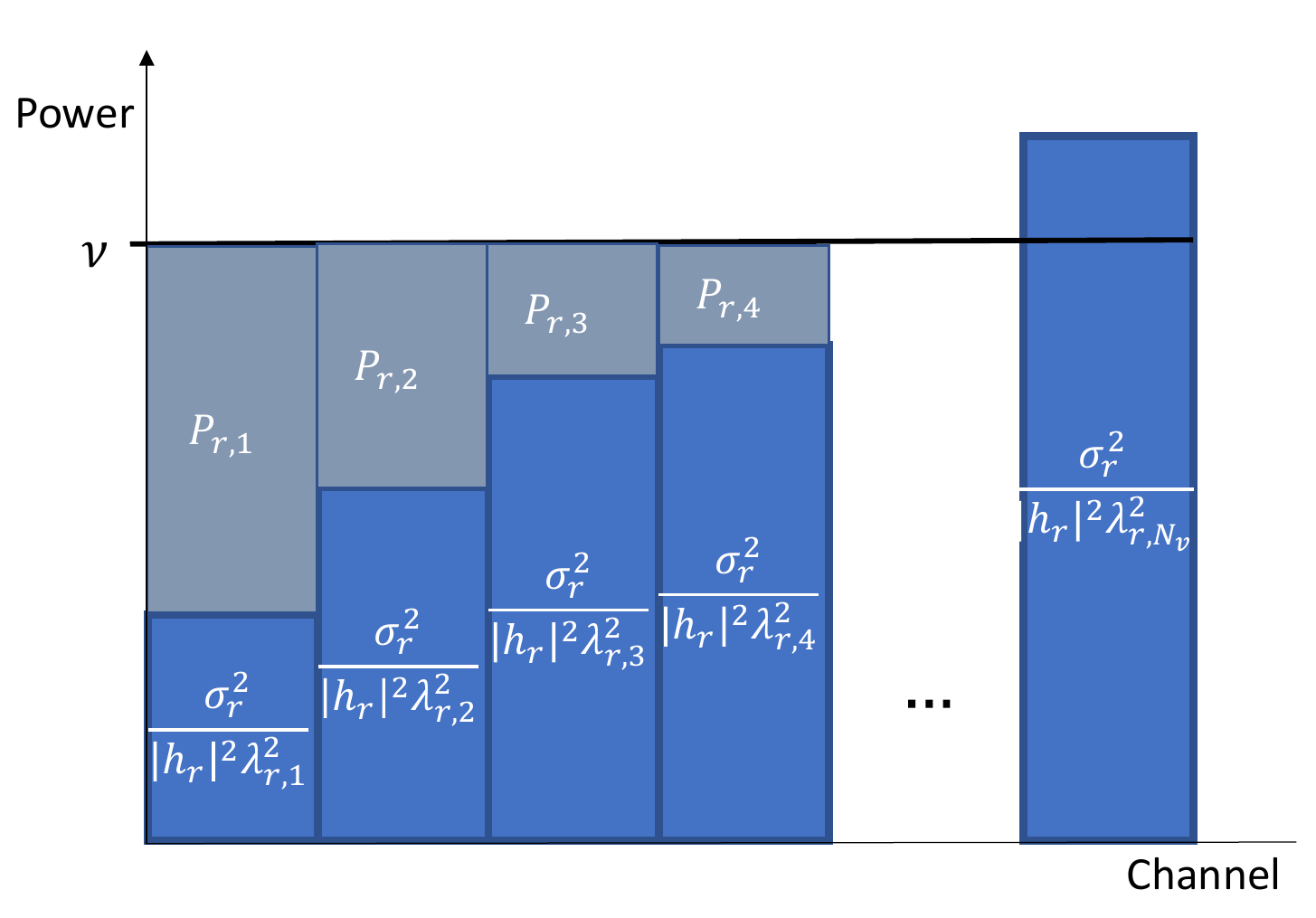}
	\caption{{\small The water-filling algorithm finds the power level $\nu$ satisfying the power constraint.}}
	\label{WF_Figure}
\end{figure}

\subsection{Achievable Rates for Two-User T-NOMA}
\label{AchievableRatesSection}
In this subsection, we review and compare the achievable rates for the two-user T-NOMA system. 
The SVD baseline decouples the superposition channel into $KN$ sub-channels, where $K=2$ for the two-user channel with the channel fading coefficients for user 1 and 2, given by $|h_r|$ for $r=1,2$. Hence, the SVD average rate per user is \cite{TimeAsynchronousNOMAforDownlinkTransmission_Ganji}:
\begin{align}
	R = \dfrac{W}{KN}\sum_{r,n}\log_2\Bigg(1 + \dfrac{P_{r,n} \lambda_{r,n} |h_{r}|^2}{ W\sigma_r^2}\Bigg).
\end{align}
To obtain the optimal SVD power allocation, we use the water-filling algorithm over the SVD sub-channels to maximize the achievable rate as shown in Fig.~\ref{WF_Figure} (cf. \cite[Ch. 9]{Info_Theory_Cover}). It is straight-forward to show that the optimal SVD power allocation is given by:
\begin{align}
&P_{r,n} = NP\dfrac{P'_{r,n}}{\sum_{r,n} P'_{r,n}},
\label{eq: SVD_power_alloc}\\ 
&P'_{r,n} = \max\Bigg(\!\dfrac{NP+\frac{\sigma_r^2}{|h_r|^2}\sum_{r,n}\frac{1}{\lambda_{r,n}^2}}{KN} - \dfrac{\sigma_r^2}{|h_r|^2\lambda_{r,n}^2}, 0\Bigg).
\label{eq: SVD_power_norm}
\end{align}
%
%

By contrast, in T-NOMA with stronger/weaker user selection, we assume the users’ channel qualities are sorted such that $|h_s|^2/\sigma_s^2>|h_w|^2/\sigma_w^2$, where subscripts $s$ and $w$ denote the stronger and weaker users, respectively. The stronger user can correctly decode their message as well as the weaker user’s message. However, the weaker user considers the stronger user's message as noise for decoding. The instantaneous achievable rates for the T-NOMA with stronger/weaker user selection are expressed by \cite{FasterthanNyquistbraodcastsignaling_Kim}:
\begin{align}
	R_{s} 
	&\leq \dfrac{1}{2T}\log_2\Bigg(1 + \dfrac{P_s|h_s|^2}{W\sigma_s^2}\Bigg), \\
	R_{w} 
	&\leq \dfrac{1}{2T}\log_2\Bigg(1 + \dfrac{P_w|h_w|^2}{P_sG_{2,1}|h_w|^2 + W\sigma_w^2} \Bigg),
\end{align}
where $P_s + P_w = P$, $W = 1/(2T)$, and $G_{2,1} \triangleq \sum_i g_{2,1}^2[i]$.

\begin{figure}[!ht]
	\centering
	\begin{subfigure}[t]{\textwidth}
\includegraphics[width=8.5cm, height=6.0cm]{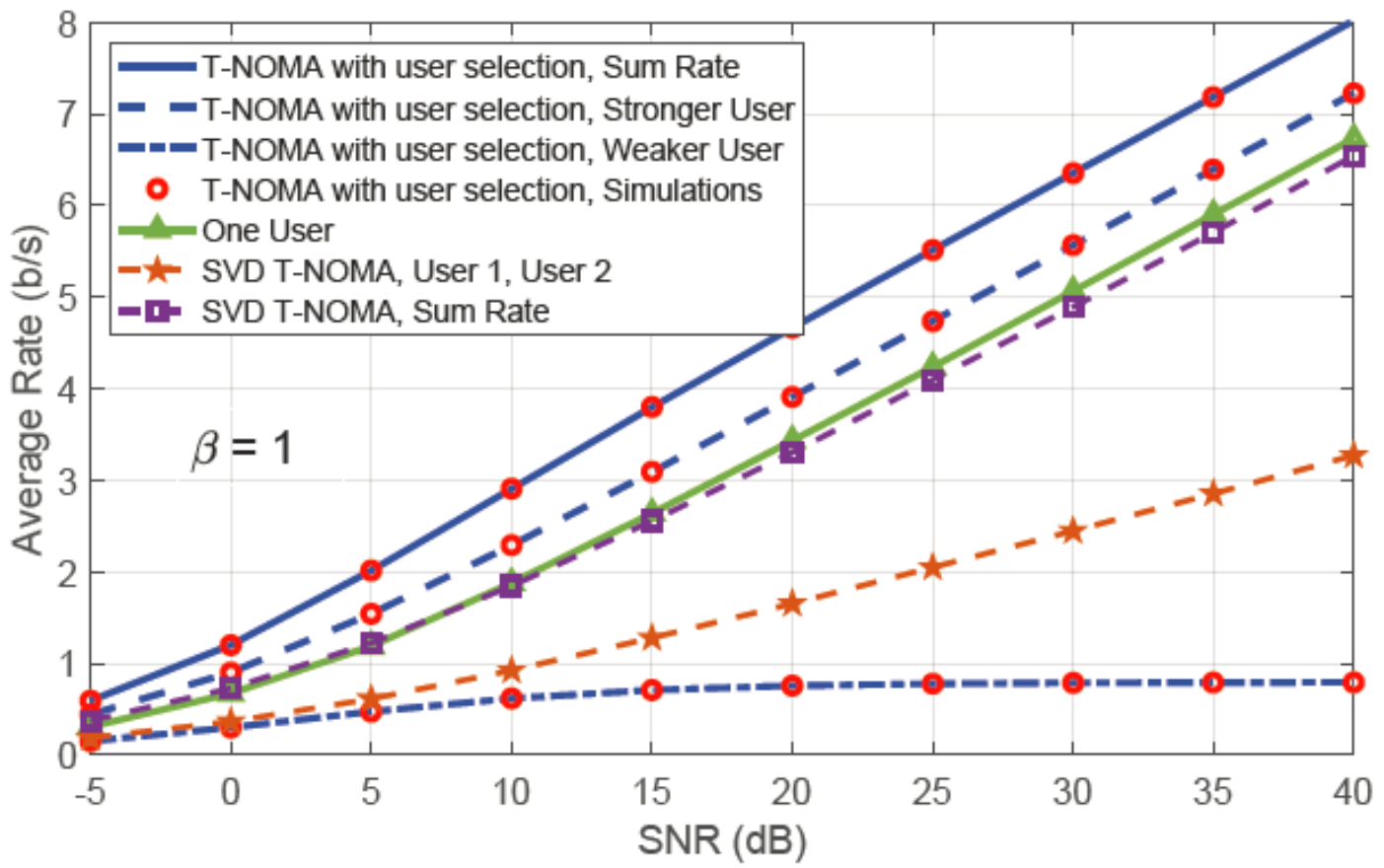}
	\end{subfigure} 
	\begin{subfigure}[t]{\textwidth}
\includegraphics[width=8.5 cm, height=6.0cm]{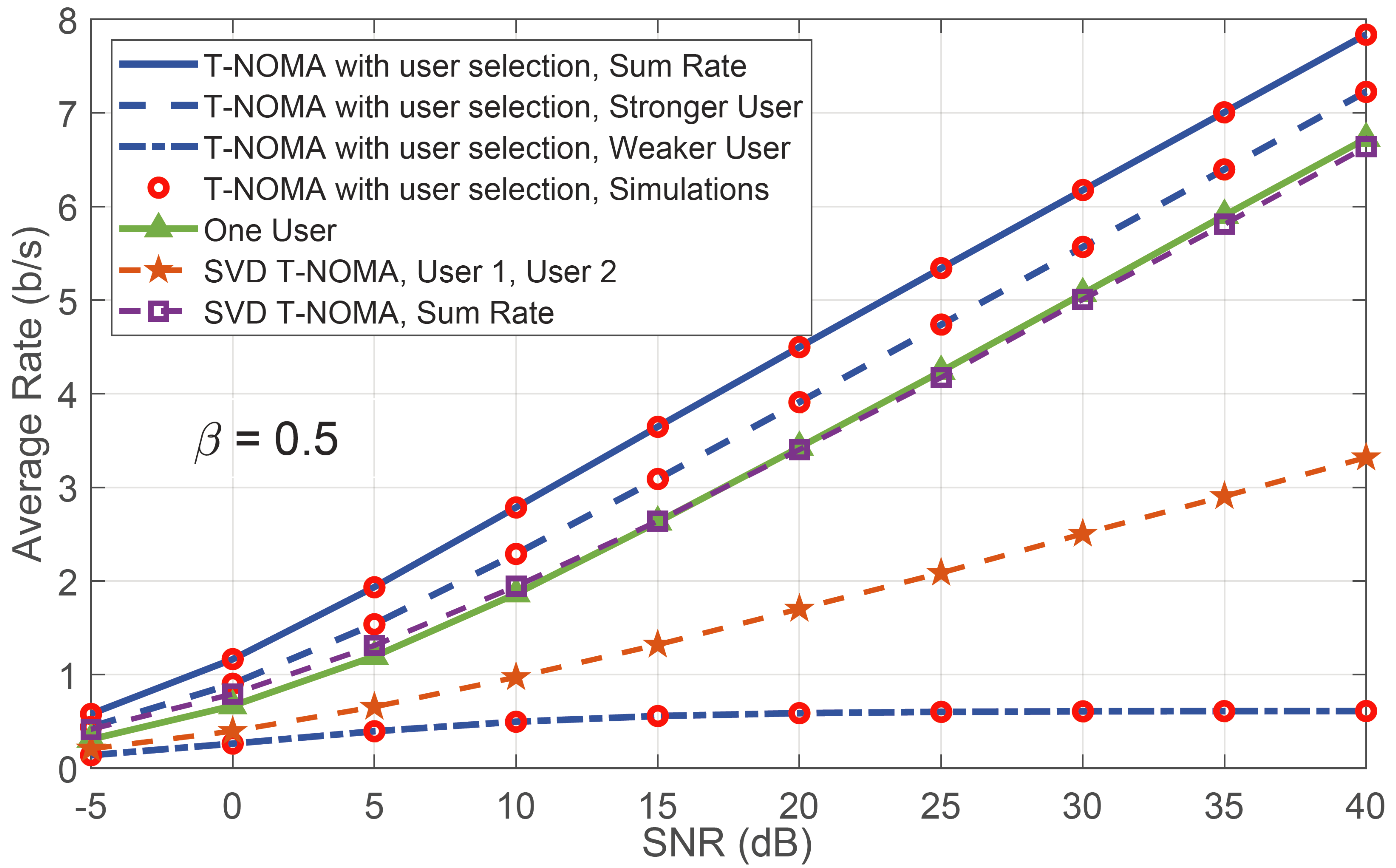}
\end{subfigure}
	\caption{{\small Achievable rates for T-NOMA on the Rayleigh fading channel.}}
\label{AchievableRatesFigure}
\end{figure}
Fig.~\ref{AchievableRatesFigure} compares the ergodic achievable rates for two-user T-NOMA for different values of the RCP roll-off factor $\beta$. The theoretical ergodic achievable rates for T-NOMA with stronger/weaker user selection are derived in Appendix \ref{AverageRateAnalysisAppendix}. We compare the theoretical rates with numerically averaged rates over many realizations of the channel fading coefficients. To compute SVD achievable rates, we perform power allocation per \eqref{eq: SVD_power_alloc} and \eqref{eq: SVD_power_norm} and average over many channel realizations.  We assume that $\sigma_1^2 = \sigma_2^2 = N_0$. The average SNR per user is defined as $P / (2N_0) $ for a fair comparison with the one-user system with power $P_1$. We set $P_1 = P_2 = 1$ for T-NOMA with stronger/weaker user selection and the one-user systems. For the SVD, we set $P=2$, $K =2$, and $N=512$ and allocate sub-channel powers according to \eqref{eq: SVD_power_norm}. Furthermore, we assume a Rayleigh fading distribution, i.e., $h_r \sim \mathcal{CN}(0,1)$. We use an offset $\tau=T/2$, the optimal offset for two-user T-NOMA \cite{ExactBERPerformanceforSymbolAsychronousTwoUserNOMA_Liu,AsynchronousTrainsmissionforMACsRateRegionAnalysisandSystemDesignforUplinkNOMA_Ganji}. 

T-NOMA with stronger/weaker user selection enables higher channel capacity than that of the single-user channel. 
But the weaker user experiences rate saturation due to the interference from the stronger user's signal. The SVD method yields lower achievable rates than stronger/weaker user selection and comparable rates to the single user. Deriving theoretical achievable rates for the AE remains an open problem. 

We conclude with a few additional remarks about the above performance results. For SVD, the $r$th user's received signal is given in \eqref{eq:SVD_yr}. For the AE, the $r$th user's received signal is
\begin{equation}
\mathbf{y}_{\!\mathrm{AE}}^{r} = f_r\left(q_rh_r \mathbf{G}\mathbf{P}_{\!\mathrm{AE}} f_e(\mathbf{x}) + q_r\mathbf{n}^r\right),
\label{eq:AE_yr}
\end{equation} 
where $f_r$ is the $r$th user's CNN decoding function, $q_r$ is the $r$th user's receiver MLP combining coefficient, $\mathbf{G}$ is the convolution matrix (see \eqref{eq:SVD_yr}) that accounts for ISI and IUI, $\mathbf{P}_{\!\mathrm{AE}}$ is the power allocation from the transmitter MLP, and $f_e$ is the transmitter's CNN encoding function. Comparing the received signals for SVD and AE, we note the following:
\begin{enumerate}[wide, labelwidth=!, labelindent=0pt]
\item Denote the achievable rates of SVD and AE by $R_{\mathrm{SVD}}$ and $R_{\mathrm{AE}}$ and that of T-NOMA with stronger/weaker user selection (where only $\mathbf{Px}$ is sent) by $R_{\text{S/W T-NOMA}}$ (also referred to as sub-FTN NOMA in \cite{FasterthanNyquistBroadcastinginGaussianChannelsAchievableRateRegionsandCoding_Kim}). Equations \eqref{eq:SVD_yr} and \eqref{eq:AE_yr} are preprocessing operations on the transmitted symbol vector $\mathbf{x}$. Hence, by the data processing inequality \cite{Info_Theory_Cover}, $R_{\mathrm{SVD}} \le R_{\text{S/W T-NOMA}}$ and $R_{\mathrm{AE}} \le R_{\text{S/W T-NOMA}}$.
\item In general, the functions considered by the AE in its training optimization include any linear processing.  In fact, in theory, a CNN is a general function approximator \cite{Heinecke2020}.
\item Let $P_{\mathrm{e,AE}}$ and $P_{\mathrm{e,SVD}}$ denote the AE and SVD BERs, respectively.  We claim that $P_{\mathrm{e,AE}} \le P_{\mathrm{e,SVD}}$. This follows because: (i) the AE is trained to minimize the CE (or the combined CE and Q-function loss), and hence minimizes the BER, whereas SVD is simply an orthogonalizing transform; and (ii) as in point 2 above, the AE (in general) includes any linear processing, and hence could be trained to replicate SVD processing.
\end{enumerate}

	\section{Conclusion}
We propose a convolutional neural network (CNN) auto-encoder (AE) for downlink time-offset with faster than Nyquist signaling non-orthogonal multiple access (T-NOMA). As in previous studies, our simulations demonstrate that SVD-based and AE-based T-NOMA satisfy the user fairness criterion, whereas T-NOMA with user selection does not. In T-NOMA, the SVD baseline's encoding and decoding complexities scale quadratically in the transmitted sequence length. In comparison, the proposed AE's encoding and decoding complexities are linear in the transmitted sequence length. Simulations demonstrate that the AE outperforms the SVD baseline by approximately 2 dB for BERs below $10^{-2}$ under perfect CSI. We have also proposed a multilayer perceptron (MLP) power allocator (MLP-PA) and an MLP transformer (MLP-T) for power allocation and channel state information (CSI) transformation that are trained using the end-to-end performance. The AE with MLP-PA outperforms the SVD baseline by about 6 dB at BERs of about $10^{-3}$. Furthermore, we have demonstrated that the CNN AE system is more resilient to timing error and imperfect CSI than the SVD baseline. 

Future research can investigate the impact of CSI errors and timing errors on the achievable rates, as well as practical channel coding schemes for the SVD and AE. SVD code rates can be optimized based on the subchannel SNR. Efficient code rate allocation for the AE remains an open challenge.
\appendices	
\section{Sufficient Statistic Samples for Matched Filter Output in T-NOMA}
\label{Sufficient_Statistics_Appendix}
Let $p(t)$ be the root raised cosine pulse. Assuming $P_{k,n} = 1$, $h_r=1$, and no noise, then $y^r(t)$ is given by
\begin{align}
	y^r(t) = &\sum_nx_1[n]p(t- T/2 - (n-1)T) \nonumber \\ + &\sum_nx_2[n]p (t - (n-1)T)
\end{align}

Then, the sufficient statistics sequence $[\mathbf{y}_1^r,\mathbf{y}_2^r]$ is obtained as follows. The sequence $\mathbf{y}_1^r$ is obtained by correlating $y^r(t)$ with $p (t - (m-1)T - T/2)$. That is,
\begin{align}
	&y_1^r[m] = \\&\int_{-\infty}^{\infty} y^r(\lambda)p(\lambda-(m-1)T-T/2)d\lambda = \\
	&\sum_nx_1[n]\int_{-\infty}^{\infty}p(\lambda -T/2 - (n-1)T) \times \ldots \\ & \qquad \qquad \qquad \times p(\lambda - (m-1)T - T/2)d\lambda + \ldots\nonumber \\
	&+\sum_nx_2[n]\int_{-\infty}^{\infty}p(\lambda - (n-1)T) \times \ldots  \nonumber \\  &\qquad \qquad \qquad \times p(\lambda - (m-1)T - T/2)d\lambda.
\end{align}

Using a change of variable $\lambda^\prime = \lambda - T/2 - (n-1)T$ for the first integral, and a change of variable $\lambda^\prime = \lambda - (n-1)T$ for the second integral, we get
\begin{align}
	y_1^r[m] =&\sum_nx_1[n]\int_{-\infty}^{\infty}p(\lambda^\prime)p(\lambda^\prime + (n-m)T)d\lambda^\prime \nonumber \\
	&+\sum_nx_2[n]\int_{-\infty}^{\infty}p(\lambda^\prime)p(\lambda^\prime+(n-m)T - T/2)d\lambda^\prime \\
	=& \sum_nx_1[n]\int_{-\infty}^{\infty}p(\lambda^\prime)p( (m-n)T -\lambda^\prime)d\lambda^\prime \nonumber \\
	&+\sum_nx_2[n]\int_{-\infty}^{\infty}p(\lambda^\prime)p((m-n)T + T/2-\lambda^\prime)d\lambda^\prime,
\end{align}
where the last equality follows by the even symmetry of $p(t)$. Thus,
\begin{align}
	y_1^r[m] = &\sum_nx_1[n] (p * p)(t)|_{t= (m-n)T} + \ldots \nonumber \\ & + \sum_n x_2[n] (p * p)(t)|_{t= (m-n)T+T/2}.
\end{align}

The sequence $\mathbf{y}_2^r$ is obtained by correlating $y^r(t)$ with $p (t - (m-1)T)$. Hence, 
\begin{align}
	&y_2^r[m] \nonumber \\
 =&\int_{-\infty}^{\infty} y^r(\lambda)p(\lambda-(m-1)T)d\lambda \\
	=&\sum_nx_1[n]\int_{-\infty}^{\infty}p(\lambda -T/2 - (n-1)T)p(\lambda - (m-1)T)d\lambda \nonumber \\
	&+\sum_nx_2[n]\int_{-\infty}^{\infty}p(\lambda - (n-1)T)p(\lambda - (m-1)T )d\lambda. \label{SecondSample}
\end{align}
Using a change of variable $\lambda^\prime = \lambda - T/2 - (n-1)T$ for the first integral, and a change of variable of $\lambda^\prime = \lambda - (n-1)T$ for the second integral, we get
\begin{align}
	&y_2^r[m] \nonumber\\ =&\sum_nx_1[n]\int_{-\infty}^{\infty}p(\lambda^\prime)p(\lambda^\prime + (n-m)T + T/2)d\lambda^\prime \nonumber \\
	&+\sum_nx_2[n]\int_{-\infty}^{\infty}p(\lambda^\prime)p(\lambda^\prime+(n-m)T)d\lambda^\prime \\
	=& \sum_nx_1[n]\int_{-\infty}^{\infty}p(\lambda^\prime)p( (m-n)T - T/2 -\lambda^\prime)d\lambda^\prime \nonumber \\
	&+\sum_nx_2[n]\int_{-\infty}^{\infty}p(\lambda^\prime)p((m-n)T -\lambda^\prime)d\lambda^\prime,
\end{align}
where the last equality follows by the even symmetry of $p(t)$.
Thus, we get 
\begin{align}
	y_2^r[m] = &\sum_nx_1[n] (p * p)(t)|_{t= (m-n)T -T/2} \nonumber \\ &+ \sum_n x_2[n] (p * p)(t)|_{t= (m-n)T}.
\end{align}	

\section{Gradients of the Q-function-loss}
\label{App A}
We show how the gradient of the Q-function loss can be computed to update the weights in the AE's last layer. Denote the estimate of the probability that a transmitted bit is $+1$ for mini-batch $l$ and bit index $n$ by $\hat{p}_{l,n}$. As a function of the final layer weights $w_j$, $\hat{p}_{l,n}$ can be written as
\begin{align}
	\hat{p}_{l,n} = \phi\Big(\sum_j w_jo^l_{n-j}\Big),
\end{align}
where $\phi(x)$, the final layer activation, can be the sigmoid function $\phi(x)=  1/(1+\exp(-x))$, and $o^l_{n-j}$ is the input to the final layer for the $n$-th bit in the $l$-th training mini-batch.

The partial derivative of the Q-function loss term $ J_{Q}$ with respect to the final layer is obtained using the chain rule as
\begin{align}
	\dfrac{\partial J_{Q}}{\partial w_j} &= \dfrac{\partial  \dfrac{1}{2}\sum_{s,l,n}Q\bigg(\kappa\dfrac{\mu_s}{\sigma_s}\bigg)}{\partial w_j}\nonumber \\ &= \dfrac{1}{4}\sum_{s,l,n} \dfrac{\partial\text{erfc}\left(\!\kappa \frac{\mu_s}{\sqrt2\sigma_s}\!\right)}{ \partial \frac{\mu_s}{\sigma_s}}	\dfrac{\partial \frac{\mu_s}{\sigma_s} }{\partial \text{LLR}^s_{l,n}} \dfrac{\partial \text{LLR}^s_{l,n} }{\partial \hat{p}_{l,n}} \dfrac{\partial \hat{p}_{l,n}}{\partial w_j}.
	\label{ChainRuleQloss_Equation}
\end{align}

We then evaluate the individual terms in \eqref{ChainRuleQloss_Equation}.
The derivative $\partial \hat{p}_{l,n} / \partial w_j$ is given by
\begin{align}
	\dfrac{\partial \hat{p}_{l,n}} {\partial w_j} =  \phi'\Big(\sum_j w_jo_{i-j}\Big)o^l_{n-j}.
\end{align}

The LLR is defined as
\begin{align}
	\text{LLR}_{l,s} &\triangleq  \log\Big(\dfrac{\Pr \{\hat{x}_{l,n} = +1\}}{\Pr \{\hat{x}_{l,n} = -1\}}\Big) = \log\Big(\dfrac{\hat{p}_{l,n}}{1-\hat{p}_{l,n}}\Big). 
\end{align}
The derivative of the LLR with respect to $\hat{p}_{l,n}$ is given by
\begin{align}
	\dfrac{\partial \text{LLR}_{l,n}}{\partial \hat{p}_{l,n}} = \dfrac{1}{\hat{p}_{l,n}(1-\hat{p}_{l,n})}. \label{Derivative_LLR_wrt_probability}
\end{align}
Note that gradient clipping is typically used to prevent large gradients in \eqref{Derivative_LLR_wrt_probability}.
From \eqref{mean_LLRs_Eq}, it is straightforward to show that
\begin{align}
	\dfrac{\partial \mu_s / \sigma_s}{\partial \text{LLR}^s_{l,n}} = \dfrac{1}{L_sN_s}\Bigg( \dfrac{1}{\sigma_s} - \dfrac{2\mu_s(\text{LLR}_{l,n}^s-\mu_s)}{\sigma_s^3}\Bigg).
\end{align}
Since  $d (\text{erfc}(ax)) / dx = -(2a/\sqrt{\pi})\exp(-(ax)^2) $, we get that
\begin{align}
	\dfrac{\partial\text{erfc}\left( \kappa \frac{\mu_s}{\sqrt2\sigma_s} \right)}{\partial \frac{\mu_s}{\sigma_s}} &= -\kappa\sqrt{\dfrac{2}{\pi}}\exp\Big(-\dfrac{1}{2}\Big(\kappa\dfrac{\mu_s}{\sigma_s}\Big)^2\Big).
\end{align}
%
\section{Average Rate analysis for T-NOMA with stronger/weaker user selection (sub-FTN-NOMA)}	
\label{AverageRateAnalysisAppendix}
%
For sub-FTN-NOMA with two users, the users' instantaneous
SNRs have been sorted such that $|h_s|^2/\sigma_s^2>|h_w|^2/\sigma_w^2$. Therefore, the user associated to $\gamma_s=|h_s|^2/(W\sigma_s^2)$ is considered as the stronger user and the user associated to $\gamma_w=|h_w|^2/(W\sigma_w^2)$ is considered as the weaker user, in which $W=1/(2T)$. We assume that the \emph{unsorted} users' SNRs $\alpha_1$ and $\alpha_2$ follow the Rayleigh fading distribution, such that, for $i=1,2$, the cumulative distribution function (CDF) of $\alpha_i \triangleq |h_i|^2/\sigma_i^2$  is given by $F_{\alpha_i}(y)=1-e^{-y/\overline{\alpha}_i}$,  where $\overline{\alpha}_i$ is the average value of ${\alpha}_i$. Accordingly, we can easily show that for the stronger user with SNR $\gamma_s=\max\left(\alpha_1,\alpha_2\right)$, its CDF expression can be expressed as $F_{\gamma_s}(y)=\prod_{i=1}^{2}F_{\alpha_i}(y)$, which can be written as $F_{\gamma_s}(y)=1-e^{-y/\overline{\alpha}_1}-e^{-y/\overline{\alpha}_2}+
e^{-y(1/\overline{\alpha}_1+1/\overline{\alpha}_2)}$. However, for the weaker user with SNR $\gamma_w=\min\left(\alpha_1,\alpha_2\right)$, we can obtain the CDF expression as $F_{\gamma_w}(y)=1-\left[1-F_{\alpha_1}(y)\right]\left[1-F_{\alpha_2}(y)\right]=
1-e^{-y(1/\overline{\alpha}_1+1/\overline{\alpha}_2)}$.

The instantaneous achievable rates for the stronger and weaker users can be expressed as \cite{TimeAsynchronousNOMAforDownlinkTransmission_Ganji}:
\begin{align}
	&R_s=W\log_2\left(1+\frac{P_s|h_s|^2}{W\sigma_s^2}\right),
	\nonumber\\
	&R_w=W\log_2\left(1+\frac{P_w|h_w|^2}{G_{21}P_s|h_w|^2+W\sigma_w^2}\right),
	\label{EQ_C2}
\end{align}
where $P_s$ and $P_w$ denote the allocated power to the stronger and weaker users, respectively, and $P=P_s+P_w$ is the total allocated power, and $G_{2,1} \triangleq \sum_i g_{2,1}^2[i]$. 
Assuming that $X_s=\gamma_s\,P_s$, where $\gamma_s=|h_s|^2/(W\sigma_s^2)$, the average rate for the stronger user $\overline{R}_s$ can be obtained by averaging $R_s$ using the probability density function (PDF) of $X_s$, $f_{X_s}(x_s)$; hence, $\overline{R}_s$ can be written as a function of $F_{X_s}(x_s)$ as:
\begin{align}
	\overline{R}_s&=W\int_0^\infty \log_2\left(1+x_s\right)f_{X_s}(x_s)\,d{x_s}
	\nonumber\\&
	=W\log_2(e)\int_0^\infty \frac{1-F_{X_s}(x_s)}{1+x_s}\,d{x_s},
	\label{EQ_C3}
\end{align}
where the second equality follows by integration by parts.
This is somewhat similar to an approach in \cite{SadjadpourNOMA} for the average rate of a NOMA system.

Since $F_{X_s}(x_s)={\rm P}(X_s\leq x_s)={\rm P}(\gamma_s\,P_s\leq x_s)=F_{\gamma_s}(x_s/P_s)$, then, $F_{X_s}(x_s)=1-e^{-x_s/(P_s\overline{\alpha}_1)}-e^{-x_s/(P_s\overline{\alpha}_2)}+
e^{-x_s(1/(P_s\overline{\alpha}_1)+1/(P_s\overline{\alpha}_2))}$. Then, \eqref{EQ_C3} becomes
\begin{align}
	\overline{R}_s&=W\log_2(e)\int_0^\infty \frac{1}{1+x_s}
	\left[e^{-\frac{x_s}{P_s\overline{\alpha}_1}}
	\right.\nonumber\\&\qquad\qquad\;\quad\left.
	+e^{-\frac{x_s}{P_s\overline{\alpha}_2}}-
	e^{-x_s(\frac{1}{P_s\overline{\alpha}_1}+\frac{1}{P_s\overline{\alpha}_2})}\right]\,d{x_s}.
	\label{EQ_C4}
\end{align}

The first integral in \eqref{EQ_C4} can be evaluated as $\int_0^\infty \frac{1}{1+x_s}e^{-x_s/(P_s\overline{\alpha}_1)}\,d{x_s}=\int_1^\infty \frac{1}{t}e^{-(t-1)/(P_s\overline{\alpha}_1)}\,d{t}$ $=e^{1/(P_s\overline{\alpha}_1)}\int_1^\infty \frac{1}{t}e^{-t/(P_s\overline{\alpha}_1)}\,d{t}=e^{1/(P_s\overline{\alpha}_1)}E_1(\frac{1}{P_s\overline{\alpha}_1})$, where we have used $\int_{1}^\infty t^{-1} e^{-\lambda t}dt=E_1(\lambda)$, for $\lambda > 0$, where $E_1(.)$ is the first-order exponential integral. Similarly, we can evaluate the remaining integral terms in \eqref{EQ_C4}, yielding
\begin{align}
	\overline{R}_s&=W\log_2(e)
	\left[e^{\frac{1}{P_s\overline{\alpha}_1}}E_1\left(\frac{1}{P_s\overline{\alpha}_1}\right)
	\right.	\nonumber\\     &\left.
	+e^{\frac{1}{P_s\overline{\alpha}_2}}E_1\left(\frac{1}{P_s\overline{\alpha}_2}\right)
	-e^{\frac{\overline{\alpha}_1+\overline{\alpha}_2}{P_s\overline{\alpha}_1\overline{\alpha}_2}}
	E_1\left({\frac{\overline{\alpha}_1+\overline{\alpha}_2}{P_s\overline{\alpha}_1\overline{\alpha}_2}}\right)
	\right],\label{EQ_C5}
\end{align}
the average rate for the stronger user in sub-FTN-NOMA.

For the weaker user, \eqref{EQ_C2} can be re-written as:
\begin{align}
	{R}_w&=W\log_2\left(1+\frac{P_w|h_w|^2}{G_{21}P_s|h_w|^2+W\sigma_w^2}\right)
	\nonumber\\	&
	=W\log_2\left(\frac{1+(G_{21}P_s+P_w)|h_w|^2/(W\sigma_w^2)}{1+\frac{G_{21}P_s|h_w|^2}{W\sigma_w^2}}\right)
	\nonumber\\
	&=W\log_2\left(1+\frac{(G_{21}P_s+P_w)|h_w|^2}{W\sigma_w^2}\right)
	\nonumber\\	
	&\quad -W\log_2\left(1+\frac{G_{21}P_s|h_w|^2}{W\sigma_w^2}\right).
	\label{EQ_C6}
\end{align}
Then, from \eqref{EQ_C6} and similarly to the steps used for the stronger user given in \eqref{EQ_C3}, the average rate for the weaker user can be
obtained by evaluating:
\begin{align}
	\overline{R}_w&=W\int_0^\infty \frac{1-F_{\gamma_w}(x_w/(G_{21}P_s+P_w))}{1+x_w}\,d{x_w}
	\nonumber\\&\quad
	-W\int_0^\infty \frac{1-F_{\gamma_w}(x_w/(G_{21}P_s))}{1+x_w}\,d{x_w}.
	\label{EQ_C7}
\end{align}
Substituting $F_{\gamma_w}(y)=1-e^{-y(1/\overline{\alpha}_1+1/\overline{\alpha}_2)}$ in above equation and evaluating the corresponding integrals by using $\int_{1}^\infty t^{-1} e^{-\lambda t}dt=E_1(\lambda)$, for $\lambda > 0$, we obtain:
\begin{align}
	\!\!\overline{R}_w&=W\log_2(e)
	\left[e^{\frac{\overline{\alpha}_1+\overline{\alpha}_2}{(G_{21}P_s+P_w)\overline{\alpha}_1\overline{\alpha}_2}}
	E_1\!\left(\!\!{\frac{\overline{\alpha}_1+\overline{\alpha}_2}{(G_{21}P_s+P_w)\overline{\alpha}_1\overline{\alpha}_2}}\!\right)
	\right.\nonumber\\&\qquad\qquad\;\quad\left.\quad
	-e^{\frac{\overline{\alpha}_1+\overline{\alpha}_2}{G_{21}P_s\overline{\alpha}_1\overline{\alpha}_2}}
	E_1\left({\frac{\overline{\alpha}_1+\overline{\alpha}_2}{G_{21}P_s\overline{\alpha}_1\overline{\alpha}_2}}\right)
	\right].
	\label{EQ_C8}
\end{align}
\section{Average BER Analysis for T-NOMA with stronger/weaker user selection}
\label{AverageBERAnalysisAppendix}
%
The average BER for T-NOMA with stronger/weaker user selection can be obtained from \cite{proakis2007digital}:
\begin{align}\label{EQ_B1}
	\overline{{\rm BER}}&=\int_0^\infty {\rm
		BER}({\gamma})\,f_{\gamma}(\gamma)\,d\gamma
	=-\int_0^\infty {\rm
		BER}^{'}({\gamma})\,F_{\gamma}(\gamma)\,d\gamma,
\end{align}
%
where ${\rm BER}^{'}({\gamma})$ is the first order derivative of ${\rm BER}(\gamma)$, the instantaneous BER of the modulation mode, given by ${\rm BER}(\gamma)=a\,Q\left(\sqrt{b\,\gamma}\right)$ \cite{proakis2007digital}, where the coefficients $\{a,b\}$ are constant values that depend on the modulation mode (e.g., for BPSK modulation $a=1, b=2$). Substituting $Q\left(x\right)=(1/\sqrt{2\pi})\,\int_{x}^{+\infty}
\exp(-t^2/2)dt$  into (\ref{EQ_B1}) simply yields:
\begin{equation}
	\label{EQ_B2} \overline{{\rm BER}} = \frac{a}{\sqrt{2\pi}}\,\int_0^\infty
	\,e^{-\frac{t^2}{2}}\,F_{\gamma}\left(\frac{t^2}{b}\right)\,dt.
\end{equation}
For the stronger user, substituting the CDF $F_{\gamma_s}(y)$ (in Appendix \ref{AverageRateAnalysisAppendix})  into (\ref{EQ_B2}) yields:
\begin{align}
	\label{EQ_B3} \overline{{\rm BER}}_s = \frac{a}{\sqrt{2\pi}}\,\int_0^\infty
	\,&e^{-\frac{t^2}{2}}\, \Bigg[1-e^{-\frac{t^2}{b\overline{\alpha}_1}} 
	\nonumber \\&
	-e^{-\frac{t^2}{b\overline{\alpha}_2}}+
	e^{-\frac{t^2}{b} \big(\frac{1}{\overline{\alpha}_1} + \frac{1}{\overline{\alpha}_2}\big)  }
	\Bigg]\,dt.
\end{align}
Using
~$\int_0^{\infty}e^{-\mu^2\,x^2}\,dx=\sqrt{\pi}/(2\mu)$ \cite[eq. (3.466.1)]{Tables_Intergral_2007} gives
\begin{align}
	\label{EQ_B4} \overline{{\rm BER}}_s=\frac{a}{2}-\frac{a}{2\sqrt{2}}
	&\Bigg[\Bigg(\frac{1}{2}+\frac{1}{b\overline{\alpha}_1}\Bigg)^{-0.5}
	+\left(\frac{1}{2}+\frac{1}{b\overline{\alpha}_2}\right)^{-0.5}
	\nonumber\\&
	-\left(\frac{1}{2}+\frac{\overline{\alpha}_1+\overline{\alpha}_2}{b\overline{\alpha}_1\overline{\alpha}_2}\right)^{-0.5}\Bigg].
\end{align}
For the weaker user, $\gamma_w=\frac{P_w|h_w|^2}{G_{21}P_s|h_w|^2+W\sigma_w^2}=\frac{P_w\gamma_w}{1+G_{21}P_s\gamma_w}$. Using \eqref{EQ_B1}, the average BER for the weaker user is
\begin{equation}
	\label{EQ_B5} \overline{{\rm BER}}_w = \int_0^\infty
	\!a\,Q\left(\sqrt{\frac{bP_w\gamma_w}{1+G_{21}P_s\gamma_w}}\,\right)\,f_{\gamma_w}\left(\gamma_w\right)\,d\gamma_w.
\end{equation}
Considering $f_{{\gamma}_w}(y)=\frac{\overline{\alpha}_1\overline{\alpha}_2}{\overline{\alpha}_1
	+\overline{\alpha}_2}e^{-y(1/\overline{\alpha}_1+1/\overline{\alpha}_2)}$, we can evaluate \eqref{EQ_B5} numerically.

	\bibliographystyle{IEEEtran}
	\bibliography{IEEEabrv,AE_NOMA_Paper_Journal_v5}

\end{document}